\documentclass[]{aa}  

\usepackage{graphicx}
\usepackage{txfonts}
\usepackage{indentfirst}
\usepackage{natbib}
\usepackage{ulem}

\begin{document}
  
  \title{Physical and radiative properties of the first core accretion shock}
  
   \author{B. Commer\c con
           \inst{1,2,3}
           ,
          	E. Audit\inst{2}
	   ,
 	 	G. Chabrier\inst{3,4}
	  	  \and
	    J.-P. Chi\`{e}ze\inst{2}
          }
   \offprints{B. Commer\c con}

 \institute{Max Planck Institute for Astronomy, K\"onigstuhl 17, 69117 Heidelberg, Germany\\
              \email{benoit@mpia-hd.mpg.de}
   \and
	 Laboratoire AIM, CEA/DSM - CNRS - Universit\'e Paris Diderot,
IRFU/SAp, 91191 Gif sur Yvette, France 
\and
   	\'Ecole Normale Sup\'erieure de Lyon, CRAL, UMR 5574 CNRS, Universit\'e de Lyon,
46 all\'ee d'Italie, 69364 Lyon Cedex 07, France
         \and 
School of Physics, University of Exeter, Exeter, UK EX4 4QL\\             }
             
   \date{Received November 26th, 2010; accepted February 14th, 2011}

  \abstract  {Radiative shocks  play  a dominant  role in star
 formation. The  accretion shocks  on  the first and
 second  Larson's  cores  involve  radiative processes  and  are  thus
 characteristic of radiative shocks. }  {In this study, we explore the formation of the
 first Larson's core and characterize the radiative
 and dynamical  properties of  the accretion shock, using both analytical  and numerical approaches. }  {We develop
both numerical radiation-hydrodynamics calculations and a  semi-analytical  model that  characterize  radiative  shocks in various
 physical conditions,  for radiating or  barotropic  fluids. Then,  we
 perform 1D spherical collapse  calculations of  the first
 Larson's core, using a grey approximation for the opacity of the material.  We consider  three  different models for radiative transfer, namely:
the  barotropic  approximation,  the flux limited  diffusion approximation and the more complete M1 model.  
We investigate the characteristic properties  of the collapse and of the first core formation. Comparison between the numerical results and our semi-analytical model
 for  radiative  shocks shows that this latter reproduces quite well the  core properties obtained with the
 numerical calculations.}
 {The accretion shock  on the first Larson core  is found to be supercritical, i.e. the
 post  and  pre-shock temperatures  are  equal, implying that all the accretion shock energy on the core is radiated away.  The  shock properties are well
 described  by the  semi-analytical model.   The flux limited  diffusion approximation is  found to agree quite well with the results based on the M1 model of radiative transfer, and is thus  appropriate to  study  the  star
 formation  process. In contrast, the  barotropic  approximation does  not
 correctly  describe  the  thermal  properties  of the  gas  during  the
 collapse.}   {We have explored and characterized the properties of radiative shocks typical of the formation of the first Larson's core during protostellar collapse,
 using both radiation-hydrodynamics numerical simulations and a semi-analytical model,
 except, for this latter, in the case of subcritical shocks in an optically thin medium. We show that a consistent treatment of radiation and hydrodynamics is mandatory to correctly handle the cooling of the gas during the core formation and thus to obtain the correct mechanical and thermal properties for this latter. We
 show  that the  flux limited  diffusion  approximation is  appropriate  to perform  star
 formation calculations  and thus allows a tractable and relatively correct treatment of radiative transfer in
 multidimensional radiation-hydrodynamics calculations.}  

\titlerunning{}
\authorrunning{B. Commer\c con et al.}
   \maketitle

\section{Introduction}

Star formation involves  a large variety of complex
physical processes.  Among them, radiative transfer appears  to be one
of the most important ones, since it governs the behaviour of the earliest phases
of the collapse and the formation of the so-called first core
\citep{Larson_1969}.  It is  well established  that the  molecular gas
experiences different thermal regimes,  from isothermal to adiabatic,
during the  earliest phase of  the collapse, via the  coupling between
gas,                 dust                and                radiation
\citep[e.g.][]{Larson_1969,Tscharnuter_1979,Masunaga_Miyama_Inutsuka_I_1998ApJ}.

Since the pioneering works of Winkler \&
Newman   and    Tscharnuter \citep{winkler1980,   tscharnuter1987},
relatively few studies have been devoted to the accretion shock on the
first Larson's core.  This shock is a
radiative  shock, i.e. radiation  can escape from  the infalling
material.  Once this  infalling gas has been shocked  and has reached an
optical depth of  $~1$, it becomes more or less adiabatic,  at a given entropy level characteristic of the first core initial energy content. This entropy content is kept roughly constant during the nearly adiabatic subsequent  stages of the collapse and the formation of the second core.   The
accretion shock on  the first core is thus  of prime importance, since
it determines the  entropy level of the following  stages of star formation, up to  the
formation of the protostar itself.

A  radiative shock is a shock that is
so strong  that it emits  radiation which in turn affects the  hydrodynamic behaviour of the flow. The
equations  of radiation-hydrodynamics  (RHD) must  thus be solved consistently to properly explore this  kind of a process. 
Radiative shocks  are common in  astrophysics, e.g. in  star formation
regions or around supernovae remnants. They are
well  studied   in  the   literature,  from  the   theoretical,
experimental    and     numerical     points     of     view
\citep[e.g.][]{Bouquet_2000,Drake_2007,Gonzalez_2009}.     The   large
variety of radiative shocks makes it difficult to try to classify them
\citep{Drake_2005,Michaut_2009}.

In  this work,  we  study radiative  shocks  from both numerically and
analytically. We focus on the properties  of the
accretion shock during the formation of  the first prestellar core.  In the first  part of the paper, we recall the main properties of
radiative shocks. Then, considering  the jump relations for a radiating
fluid,  we focus  on some  particular  kinds of  radiative shocks,  with
upstream   and  downstream   material  characterized by   different   optical  depth
properties. We  also consider the  case of a  shock taking place  in a
barotropic  material, since the barotropic  approximation  is the  simplest
way to characterize the thermal evolution of the  gas during the
collapse \citep[e.g.][]{Commercon_2008}.

In  the second  part  of this  work,  we study  the impact of  various
models of  radiative   transfer   on  the   protostellar
collapse.  Radiation hydrodynamics  plays here  a crucial  role, for
instance by  evacuating  the compressional  energy, leading  to a nearly isothermal
free-fall collapse phase. Radiation transfer also
has  a dramatic  impact on  the  accretion shock,  with the  infalling gas kinetic
energy  being  either  converted into  internal
energy  in the  static adiabatic  core  or radiated  away. This  ratio
between the energy accreted  onto the star and the one radiated  away is a key quantity, as  it ultimately determines the  entropy content of
the forming  star.  Using a 1D  spherical code, we  compare results of
calculations done with a barotropic  EOS, a flux  limited diffusion approximation, and a M1 model for radiative transfer, assuming grey opacity
for the infalling material. 1D calculations retain the important virtue of
allowing  detailed and complex physical processes to be considered in dense core collapse calculations, which is not the case  in a  multidimensional
approach. In consequence, a 1D approach enables us to characterize the impact of these various
physical  processes on the results, allowing a better characterization of the validity of simplified
multidimensional studies  \citep[e.g.][]{Commercon_2010L,Commercon_2011}.

This paper is  organized as follows.  In the  first section, we recall
the basic  properties of radiative shocks  and develop semi-analytical
models that  can be  applied in some  particular cases for  a radiating
fluid and a barotropic fluid.   In Sect. 3, we detail the numerical
method  and the  physics inputs used in our calculations. In  Sect. \ref{results},  we
derive the first Larson's core properties from the numerical calculations, using various approximations
for the  radiative transfer, and compare the results with the ones
obtained              by
\cite{Masunaga_Miyama_Inutsuka_I_1998ApJ}. We also present an  original semi-analytical
model  that allows a simple interpretation of the numerical  results. Eventually,
Sect. 5 summarizes our main results.

\section{Radiative shock - A semi-analytic model\label{rshock_theory}}

\subsection{A qualitative picture of radiative shocks}
%
Depending on the  shock's strength,
radiative shocks belong to two  groups: the {\it subcritical} shocks and the {\it
supercritical}  shocks. 
As  the  strength   of  a  shock  increases,  the  postshock
temperature $T_2$  rises, producing a radiative flux  of order $\sigma
T_2^4$ that increases very  rapidly. This flux penetrates the upstream
material and preheats this latter to a temperature $T_{-}$ immediately ahead of
the shock  front (radiative precursor) that  is proportional to  the incident  flux. $T_{-}$
increases  rapidly  with  the  shock strength  and  eventually  becomes equal to
$T_2$.  A  shock with  $T_{-}  < T_2$  is  called  a {\it  subcritical
shock}.  Because the material  entering   the  shock  is  preheated,  the
postshock temperature  $T_{+}$ exceeds its asymptotic equilibrium value
$T_{2}$, and  decays downstream  as the  material  cools by
emitting photons that propagate across the shock (see Fig. 1 for various shock characteristics).

For stringer shocks, the preheating
becomes so important that the  preshock temperature equals  the postshock
equilibrium temperature  $T_2$. The shock velocity  at which the  postshock
and  preshock temperatures are  equal, defines  the {\it critical  shock}. For
higher  shock velocities, $T_-$  cannot exceed  $T_2$, the  excess
energy forces the radiative precursor further into the upstream region
with a temperature close to $T_2$.  
Pre-  and post-shock  temperatures are  equal, the
supercritical shock is thus isothermal. Radiation and matter are still
out of equilibrium in some part of the precursor, but come to equilibrium
when  temperature approaches  $T_2$.  The upstream  kinetic energy  is radiated at the  shock.

 Most  of  the  early  work  on  radiative shocks  has  been described  in
\cite{Zeldovich_book} and \cite{Mihalas_book}, where readers can find the basic equations of  the front structure,  radiative precursor
extension, etc...

\begin{figure*}[htb]
  \centering 
  \includegraphics[width=14cm,height=12cm]{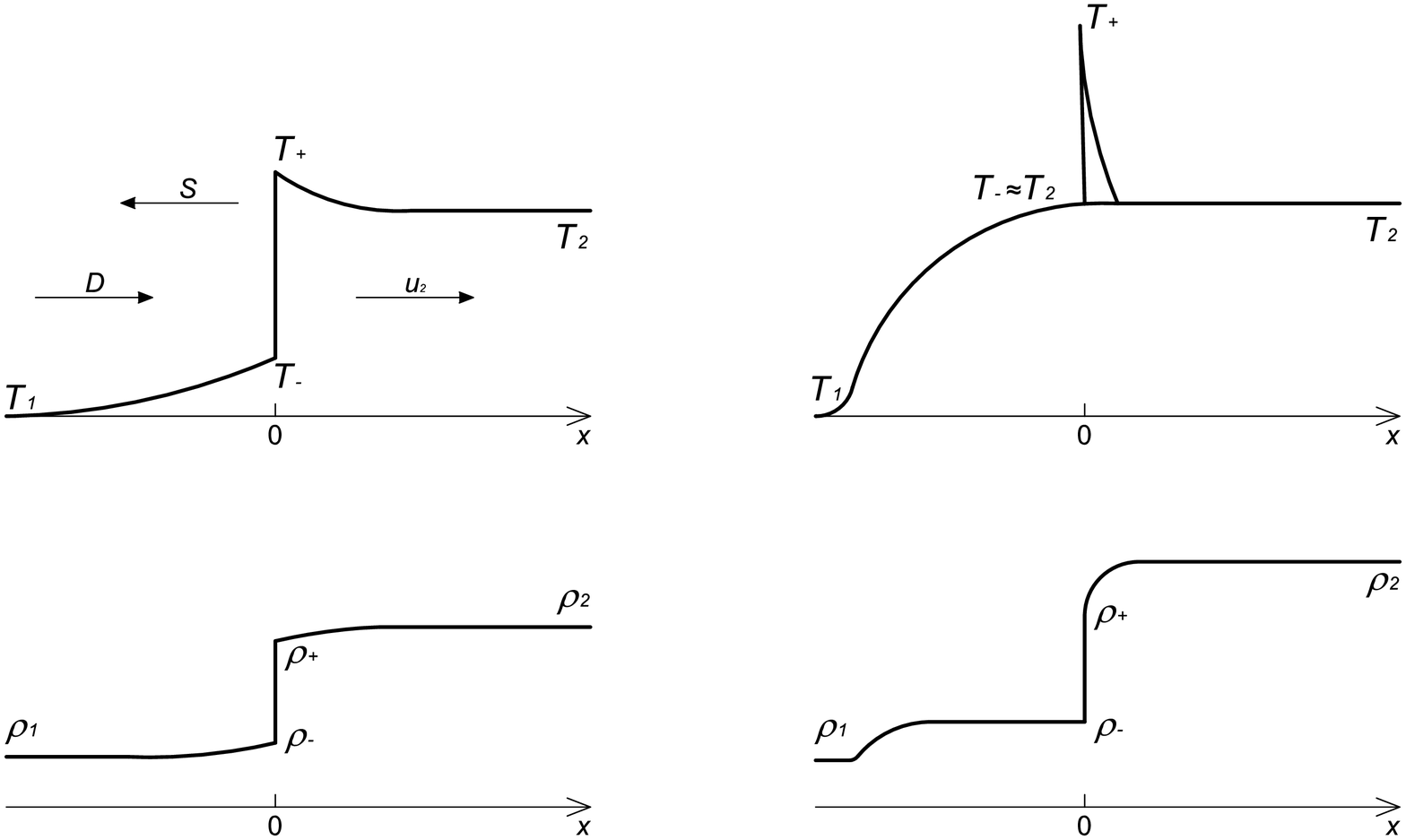}
  \caption{Temperature and density profiles in a subcritical shock ({\it left}) and a supercritical shock ({\it right}). Adapted from \cite{Zeldovich_book}. }
\label{Rshock_illu}
\end{figure*}

\subsection{Jump relations for a radiating material}

Consider the  jump relations (Rankine  Hugoniot) for a  radiating flow
\citep[see][]{Mihalas_book,Zeldovich_book,Drake_2007}.
Conservation of mass, momentum and energy yield
\begin{eqnarray}
\rho_1 u_1 & = & \rho_2 u_2 \equiv \dot{m}, \label{rh1}\\
\rho_1 u_1^2+P_1 + P_\mathrm{r1}  & = & \rho_2 u_2^2 + P_2 + P_\mathrm{r2} ,\label{rh2}\\
\dot{m} \left(h_1 +\rho_1 u_1^2\right) +F_\mathrm{r1}+u_1\left(E_\mathrm{r1}+P_\mathrm{r1} \right) & = & \nonumber \\
 \dot{m} \left(h_2 +\rho_2 u_2^2\right) &+&F_\mathrm{r2}+u_2\left(E_\mathrm{r2}+P_\mathrm{r2} \right),\label{rh3}
\end{eqnarray}
where subscripts ``1'' and  ``2'' denote respectively the upstream and
downstream  states, all  radiation quantities are estimated  in the
comoving  frame.   $h$  corresponds   to  the  gas  specific  enthalpy
($h=\epsilon  + P/\rho$,  with $e=\rho  \epsilon =  P/(\gamma-1)$) and
$\dot{m}$  is the  mass flux  through  the shock.  In comparison with  the
hydrodynamical  case,  the pressure  is  now  the  total  pressure, i.e. the gas plus radiation contributions, 
$P+P_\mathrm{r}$, and the specific enthalpy is the total specific enthalpy,
$\epsilon+(P+P_\mathrm{r}+E_\mathrm{r})/\rho$.  The radiation  energy $E_\mathrm{r}$ and
the radiation  pressure $P_\mathrm{r}$ are  important only  at high  temperatures  or low
densities, whereas the radiation flux, $F_\mathrm{r}$,  plays a fundamental role
in all radiative shocks.

Contrarily to the hydrodynamical  case, the system of equations (\ref{rh1}),
(\ref{rh2}) and  (\ref{rh3}) can not  be solved explicitly. We  need to
make some  assumptions on both the  upstream and downstream  materials. In
what follows, we distinguish  two cases: i) the upstream and downstream materials are opaque, ii) the upstream material is  optically
thin and the downstream material remains opaque.

\subsubsection{Radiative shock in an optically thick medium}

This     is     the     most     studied  case
\citep[e.g.][]{Mihalas_book,Drake_2007}. It occurs, for instance, in shocks in
stellar interiors, where matter is both dense and hot.  At a sufficiently
large distance from the front, matter and radiation are in equilibrium
and   both  the   downstream   and  upstream   materials  are   opaque
($F_\mathrm{r1}=F_\mathrm{r2}=0$).  Any radiation  crossing  the front
from the  hot downstream material  into the cool upstream  material is
reabsorbed  in the  radiative precursor,  into which  it  propagates by
diffusion. Outside this diffusion layer, the flux vanishes and we have

\begin{equation}
\small
\dot{m} \left(h_1 +\rho_1 u_1^2\right)+u_1\left(E_\mathrm{r1}+P_\mathrm{r1} \right)  = 
\dot{m} \left(h_2 +\rho_2 u_2^2\right)+u_2\left(E_\mathrm{r2}+P_\mathrm{r2} \right).
\end{equation}

Since matter and radiation are  in equilibrium and the material is opaque,
we  have  $E_\mathrm{r}=3P_\mathrm{r}=a_\mathrm{R}T^4$.  Defining  the
compression ratio $r=\rho_2/\rho_1=u_1/u_2$,  we can rewrite equations
(\ref{rh2}) and ({\ref{rh3}) in the non-dimensional form
\begin{equation}
\gamma \mathcal{M}_1^2\left(\frac{r-1}{r} \right)=\left(\Pi -1\right) +\alpha_1\left( \left(\frac{\Pi}{r}\right)^4-1\right),\label{opac1}
\end{equation}
and
\begin{equation}
\frac{\gamma}{2} \mathcal{M}_1^2\left(\frac{r^2-1}{r^2} \right)=\frac{\gamma}{\gamma-1}\left(\frac{\Pi}{r} -1\right) +4\alpha_1\left( \frac{\Pi^4}{r^5}-1\right),\label{opac2}
\end{equation}
where     $\Pi=P_2/P_1$,    $\alpha_1=(1/3)    a_\mathrm{R}T_1^4/P_1$ and
$\mathcal{M}_1$      is     the     hydrodynamic      Mach     number,
$\mathcal{M}_1=u_1/c_{s1}=u_1/(\gamma  P_1/\rho_1)^{1/2}$. The coupled  equations (\ref{opac1})
and (\ref{opac2})  are solved numerically  to get the variations  of $\Pi$
and $r$ as function of $\mathcal{M}_1$ and $\alpha_1$.

The compression ratio  rises from $r=4$ for a  pure hydrodynamic shock
with  $\gamma=5/3$  to  $r=7$,   which  corresponds  to  the  limiting
compression ratio for  a gas with $\gamma=4/3$, i.e  pure radiation, a
gas made of photons. The stronger the shock, the greater the radiative
effects, even for a very  small initial ratio of radiative pressure to
gas  pressure. When  the  upstream radiative  pressure increases,  the
shock becomes immediately radiative  since radiative pressure increases as
$T^4$,   whereas   the   gas   pressure  increases only  linearly   with
temperature. As shown in \cite{Mihalas_book}, for  a strong  radiating
shock,  since the compression  ratio is  fixed, the  temperature ratio
increases   only   as   $\mathcal{M}_1^{1/2}$,   whereas   it   rises   as
$\mathcal{M}_1^2$   in   a  non-radiating   shock.   Note  that  if
$P_\mathrm{r}/P  < 10^{-5}$, the  non-radiating fluid  approximation becomes
valid for an upstream Mach number $\mathcal{M}_1 < 10$.

\subsubsection{Radiative shock with an optically thin upstream material\label{model}}

Suppose  now  that the  shock  is  propagating  in an  optically  thin
material with  an opaque  downstream region, as  in the case  of star
formation  \citep[e.g.,  ][]{Calvet_1998}.    
In  the  low  mass  star
formation context, the radiative  energy and pressure can be neglected
compared to the gas internal energy for the first core accretion shock.

Let us consider the material and radiative quantities at the discontinuity, outside the spike in gas temperature (i.e., between regions with subscripts "2" and "-"). We have
\begin{eqnarray}
\centering
\rho_- u_- & = & \rho_2 u_2 \equiv \dot{m}, \label{rh12}\\
\rho_- u_-^2+P_-  & = & \rho_2 u_2^2 + P_2  ,\label{rh22}\\
\dot{m} \left(h_- +\rho_- u_-^2\right)  & = &
\dot{m} \left(h_2 +\rho_2 u_2^2\right) +\Delta F_\mathrm{r},\label{rh32}
\end{eqnarray}
where $\Delta  F_\mathrm{r}= F_\mathrm{r2}-F_\mathrm{r-}$. We consider
the case of a non zero net flux $|F_\mathrm{r2}-F_\mathrm{r-}|$ across the shock.  
The jump relations, derived 
from   the  conservation   equations   (\ref{rh1}),  (\ref{rh2})   and
(\ref{rh3}), read

\begin{equation}
\frac{\rho_2}{\rho_-}=\frac{\left[(\gamma +1)P_2+(\gamma-1)P_- \right]u_2 + 2(\gamma - 1)\Delta F_\mathrm{r} }
{\left[(\gamma +1)P_-+(\gamma-1)P_2\right]u_2} \label{djump},
\end{equation}
and

\begin{equation}
\frac{T_2}{T_-}=\left(\frac{P_2}{P_-}\right)\frac{\left[(\gamma +1)P_-+(\gamma-1)P_2 \right]u_- - 2(\gamma - 1)\Delta F_\mathrm{r} }
{\left[(\gamma +1)P_2+(\gamma-1)P_-\right]u_-}. \label{tjump}
\end{equation}
These  two relations show  that radiative energy  transport (i.e. the
radiative flux)  across a shock  can significantly alter  the density,
temperature  and  velocity  profiles of the flow.   Both upstream  and  downstream
materials are affected  over distances which depend on the material opacity.  As mentioned before,  the structure of a radiative shock
is  as follows:  the  upstream  material is  preheated  by a  radiation
precursor while the downstream material is cooled by radiative losses.

As for  the previous  opaque case for which we use an iterative solver,  the downstream  quantities can not be derived analytically from the conservation relations for given upstream conditions.  The
radiative flux has to be known  and the result depends on the upstream
flow.  

The simplest  case to  study  is a  supercritical shock,  where
$T_-=T_2$ and  then $h_-=h_2$. Using the  same adimensional parameters
as for the opaque case, we have
\begin{equation}
\gamma \mathcal{M}_-^2\left(\frac{r-1}{r} \right)=\Pi -1 ,
\end{equation}
and
\begin{equation}
\frac{\gamma}{2} \mathcal{M}_-^2\left(\frac{r^2-1}{r^2} \right)=\frac{\gamma}{\gamma-1}\left(\frac{\Pi}{r} -1\right) + \frac{\Delta F_\mathrm{r}}{P_- u_-}.
\end{equation}
Note that since the shock is isothermal, $r=\Pi$ and 
\begin{equation}
\Pi=\gamma \mathcal{M}_-^2.
\label{rPi}
\end{equation}
Eventually, the radiative flux discontinuity, normalized to the upstream kinetic energy, is given by:
\begin{equation}
X=\frac{\Delta F_\mathrm{r}}{0.5\rho_1 u_-^3} = \frac{\gamma^2 \mathcal{M}_-^4 -1}{\gamma^2 \mathcal{M}_-^4},
\label{XX}
\end{equation}
where $X$ thus represents the amount of incident kinetic energy radiated away at the shock front.

\begin{figure}[t]
  \centering
  \includegraphics[width=7cm,height=5.5cm]{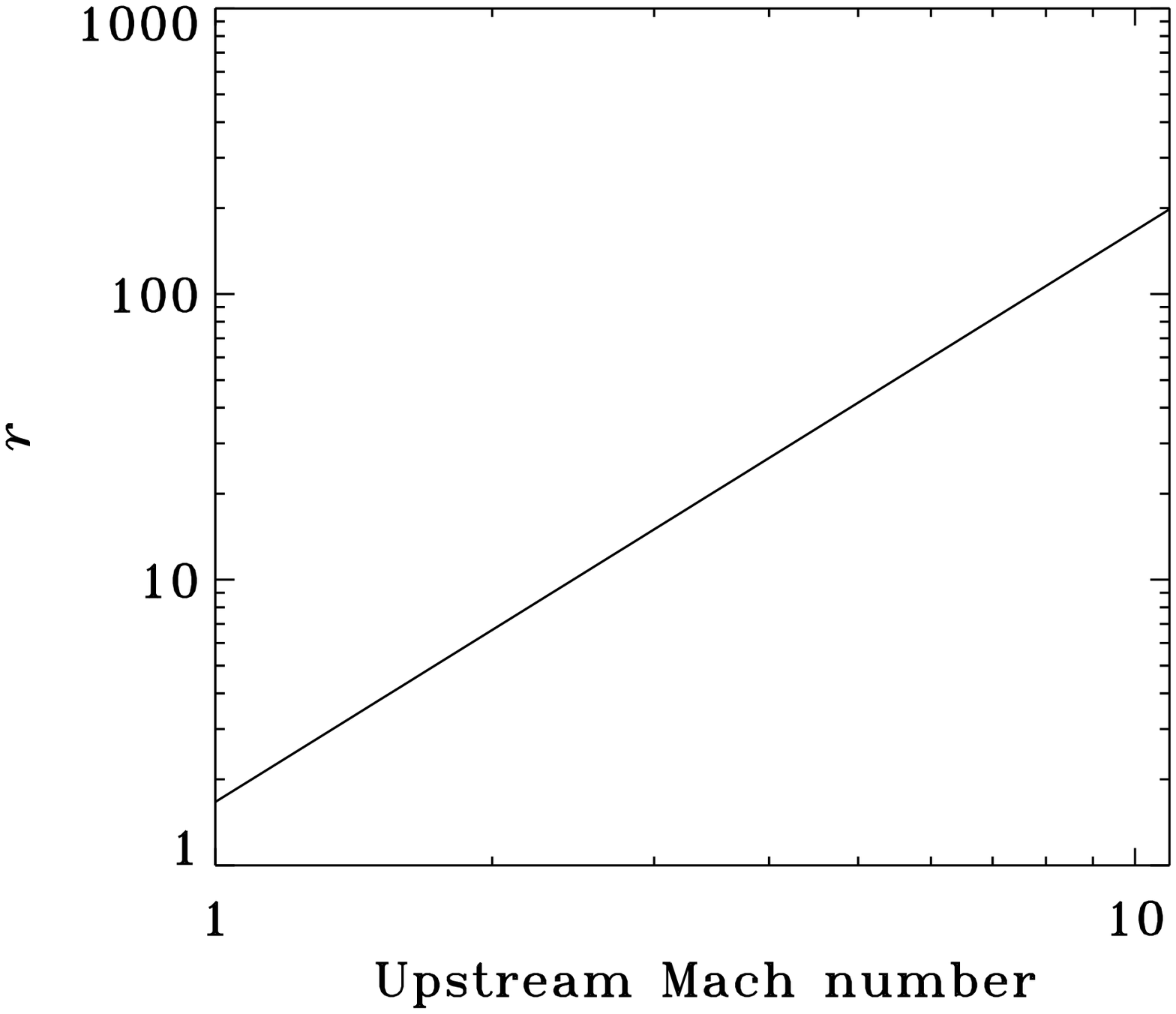}
  \includegraphics[width=7cm,height=5.5cm]{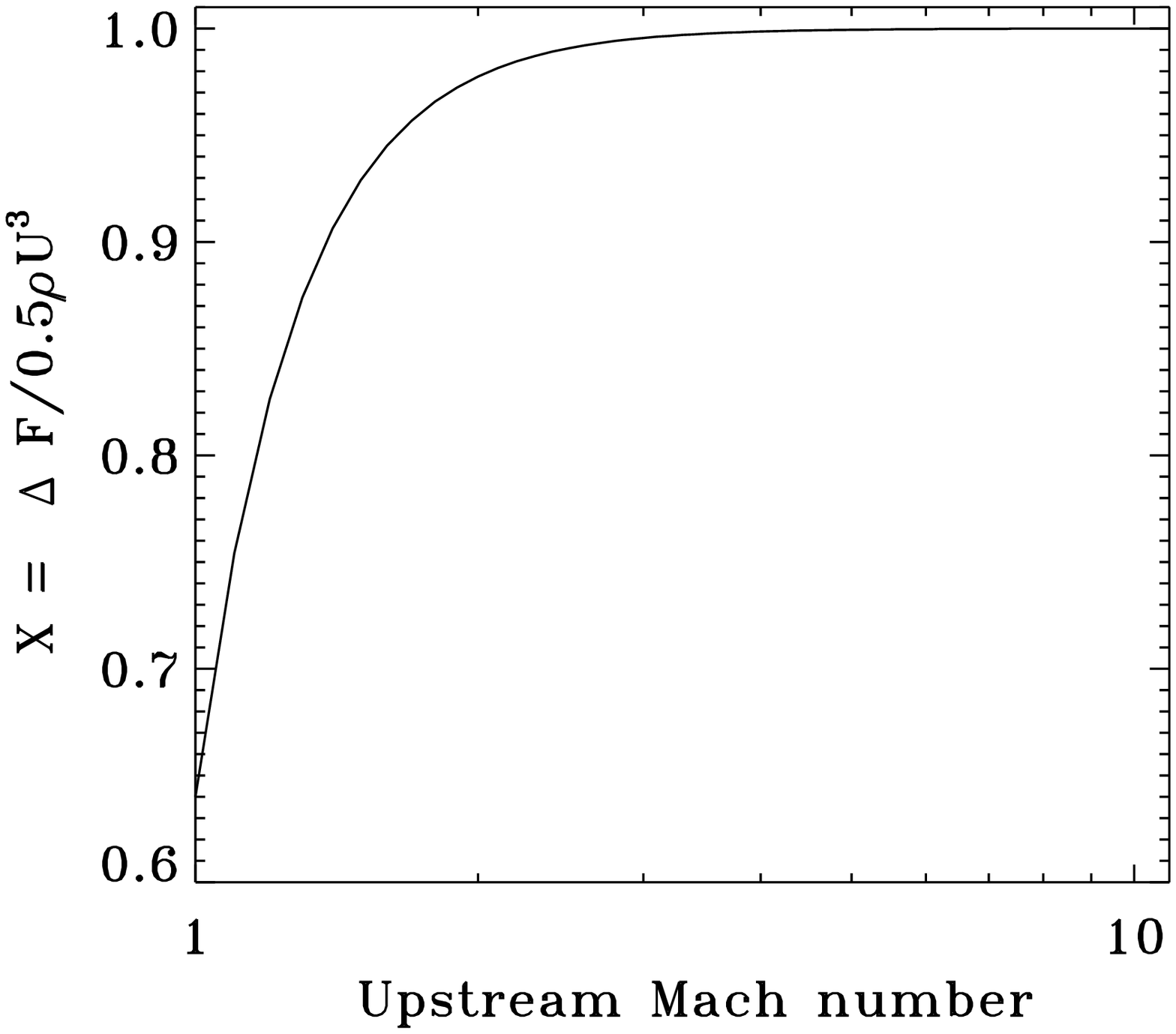}
  \caption{Compression ratio $r$ and fraction of kinetic energy radiated away at the shock, $X$, as a function of the upstream Mach number for a supercritical shock, in the case of an optically thin upstream material.}
\label{PX_fM}
\end{figure}

Figure \ref{PX_fM}  shows the evolution of the adimensional compression ratio, $r$, and flux discontinuity,
$X$, as a function of the upstream Mach  number for a  supercritical shock. As
seen in the figure, the compression ratio does not saturate, in contrast to
the  case of  an opaque  material.  As shown by  equations  (\ref{djump}) and
(\ref{tjump}), a non-zero flux across  the front,
with  $F_\mathrm{r2}>  F_\mathrm{r-}$,  increases  the  density  jump
whereas it decreases the temperature jump.  In order to have an isothermal shock
at low Mach number, $\mathcal{M}< 2$, the downstream velocity is determinant since in that case
the upstream kinetic energy  is not entirely radiated away.   In protostellar
collapse calculations, the characteristic Mach number at the
first core accretion  shock is $\mathcal{M} \sim 3$, implying that almost all the
kinetic energy ($> 99 \%$)  is radiated away at this stage.

Including  radiative  energy  and radiative pressure in  equations  (\ref{rh22})  and
(\ref{rh32}) does  not affect the dependence of  the compression ratio
upon the upstream Mach number.   On the other hand, the dependence of the amount of kinetic
energy radiated away becomes stronger:
\begin{equation}
X=\frac{\Delta F_\mathrm{r}}{0.5\rho_- u_-^3} = \frac{\gamma^2 \mathcal{M}_-^4 -1}{\gamma^2 \mathcal{M}_-^4}(1+4\alpha_1).
\end{equation}

The same analysis  cannot be carried out for  a subcritical shock with
an optically  thin upstream medium,  since, in that case, no constraint allows  us to
close the system of equations  (\ref{rh22}) and  (\ref{rh32}). One  needs a
prescription  on the  radiation  flux or  luminosity  in the  upstream
material to close the system.

\subsection{Super- or sub-critical shock ?}

In this section, we characterize in which regime the radiative shock takes place, depending on the upstream and downstream material properties.

\subsubsection{Opaque material}
In  an  opaque  material,  the  simplest  case is the one of  a
supercritical shock, in which the preshock gas ahead of the discontinuity
is  heated up  to  the postshock  temperature (see \S2.1.2).   The radiative  energy
absorbed in  the upstream region  is used only to raise  the gas
temperature \citep[][p. 536]{Zeldovich_book}.
Assuming, for sake of simplicity, that the gas is  neither compressed nor
slowed down, that the radiative pressure is negligible compared to the the gas pressure (valid for early low mass star formation stages), and that the upstream internal energy is negligible far from the shock (region with subscript "1"), the  equation of energy reads
\begin{equation}
-F_\mathrm{r}=u_-\rho_-\epsilon(T_-),
\end{equation}
with $\epsilon(T)\equiv \epsilon=\frac{1}{\gamma-1}\frac{k_\mathrm{B}T}{\mu m_\mathrm{H}}$.   
If we evaluate the flux just at the discontinuity, we find the maximum
preheating temperature T$_-$ from  
\begin{equation}
|F_\mathrm{r}|\sim \sigma T_2^4=u_-\rho_-\epsilon(T_-).
\end{equation}  

We can now determine the critical temperature $T_\mathrm{cr}$ at which
$T_-$ equals $T_2$
\begin{equation}
\sigma T_\mathrm{cr}^4=u_-\rho_-\epsilon(T_\mathrm{cr}),
\end{equation}  
so that
\begin{equation}
T_\mathrm{cr}=\left(\frac{u_-\rho_- k_\mathrm{B}}{(\gamma - 1)\mu m_\mathrm{H} \sigma}\right)^{1/3},
\label{Tcr}
\end{equation}  
which defines the supercritical shock condition.

\subsubsection{Optically thin upstream material}

In the case of an optically thin upstream material, the aforederived
criterion for a supercritical shock is simply derived by assuming that  all the  upstream kinetic
energy is radiated away at the shock, which yields

\begin{equation}
T_\mathrm{cr}=\left( \frac{0.5\rho_- u_-^3}{\sigma}\right)^{1/4}.
\label{Tcr2}
\end{equation}

\subsection{Estimate of the preshock temperature\label{sec42}}

In  this  section, we  calculate  the  preshock temperature  as a
function of the upstream quantities, for various natures of the shock.

\subsubsection{Supercritical shock with an optically thin or thick upstream material}

This is the simplest case  since the results are independent of the
optical depth of the upstream material.
Once the upstream  density is known, the velocity  and the temperature
are set. When the upstream  gas is not compressed ($u_1=u_-$) and for a
strong shock  ($\mathcal{M}>2$ so that $X\sim  1$), it is  easy to get
the  shock temperature $T_\mathrm{s}=T_-=T_2$  since all  the upstream
kinetic energy is radiated away:
\begin{equation}
T_\mathrm{s}=\left( \frac{0.5\rho_1 u_1^3}{\sigma}\right)^{1/4}.
\label{Tp_super}
\end{equation}

\subsubsection{Subcritical shock with an optically thick upstream material}
In  that  case,  the  shock  temperature  is  fixed  by  the  upstream
velocity.  The  equilibrium  postshock
temperature is ( \cite{Mihalas_book}):
\begin{equation}
T_2=\frac{2(\gamma-1)u_-^2}{\mathcal{R}(\gamma+1)^2},
\end{equation}
with $\mathcal{R}=k_\mathrm{B}/\mu m_\mathrm{H}$. The preshock temperature $T_-$ is estimated as
\begin{equation}
\frac{\dot{m}\mathcal{R}T_{-}}{\gamma-1}={2\sigma T^4_2}{\sqrt{3}},
\end{equation}
which  indicates that  at  any location ahead of  the shock,  all the
radiative energy going across this location is absorbed and heats up the gas.

\subsection{Jump relations for a barotropic gas}

A widely used approximation to handle radiative transfer during the first stages
of  star  formation  is  the  barotropic  approximation
\citep[e.g.   ][]{Commercon_2008}.   For  a barotropic  material,  the
temperature and pressure depend only  on the density. The total energy
is thus not conserved. In this case, the jump relations are simply
\begin{eqnarray}
\centering
\rho_1 u_1 & = & \rho_2 u_2 \equiv \dot{m}, \label{rh12b}\\
\rho_1 u_1^2+P_1  & = & \rho_2 u_2^2 + P_2  ,\label{rh22b}
\end{eqnarray}
with the barotropic equation of state (EOS) as a closure relation
\begin{equation}
P \propto \rho\left[1+\left(\frac{\rho}{\rho_\mathrm{ad}}\right)^{\gamma-1}\right],
\label{baro1}
\end{equation}
where  $\rho_\mathrm{ad}$ is  the critical  density at  which  the gas
becomes adiabatic (see Sect. \ref{EOS_barotrop}). Using the same adimensional parameters as previously, we get

\begin{equation}
\gamma \mathcal{M}_1^ 2\left( \frac{r-1}{r}\right) = \Pi -1.
\end{equation}
Using the barotropic EOS yields
\begin{equation}
\Pi=\frac{P_2}{P_1}=\frac{\rho_2}{\rho_1}\frac{1+r_2^{(\gamma-1)}}{1+r_1^{(\gamma-1)}}= r\frac{1+r_1^{(\gamma-1)}r^{(\gamma-1)}}{1+r_1^{(\gamma-1)}},
\end{equation}
where $r_1=\rho_1/\rho_\mathrm{ad}$ and $r_2=\rho_2/\rho_\mathrm{ad}$. Eventually, we get $r$ by solving
\begin{equation}
1+r_1^{(\gamma-1)}\gamma \mathcal{M}_1^2\frac{r-1}{r} = (r-1) + r_1^{\gamma-1}(r^ \gamma -1 ).
\label{tosolve_baro}
\end{equation}

One can also derive an equivalent luminosity, assuming a shock in a perfect gas with the same jump properties. 
From equation (\ref{rh32}), we have
\begin{equation}
X=\frac{\Delta F_\mathrm{r}}{0.5\rho_1 u_1^3} = \frac{2}{\gamma-1}\frac{1}{\mathcal{M}_1^2}\left(1-\frac{\Pi}{r} \right) + \left(1-\frac{1}{r^2}\right),
\end{equation}
where $r$ and $\Pi$ are given by (\ref{tosolve_baro}). 

\begin{figure}[thb]
  \centering
  \includegraphics[width=7cm,height=5.5cm]{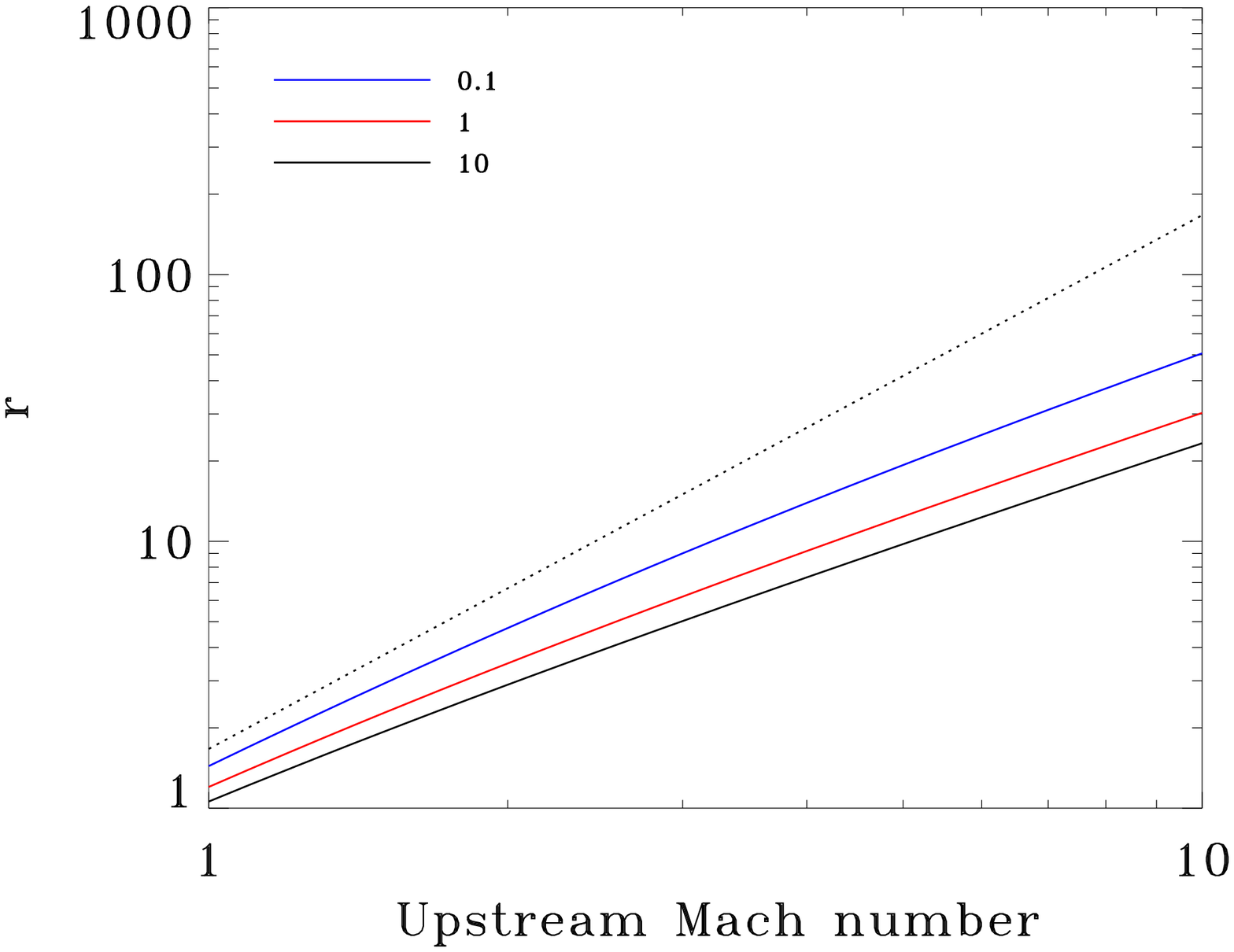}
  \includegraphics[width=7cm,height=5.5cm]{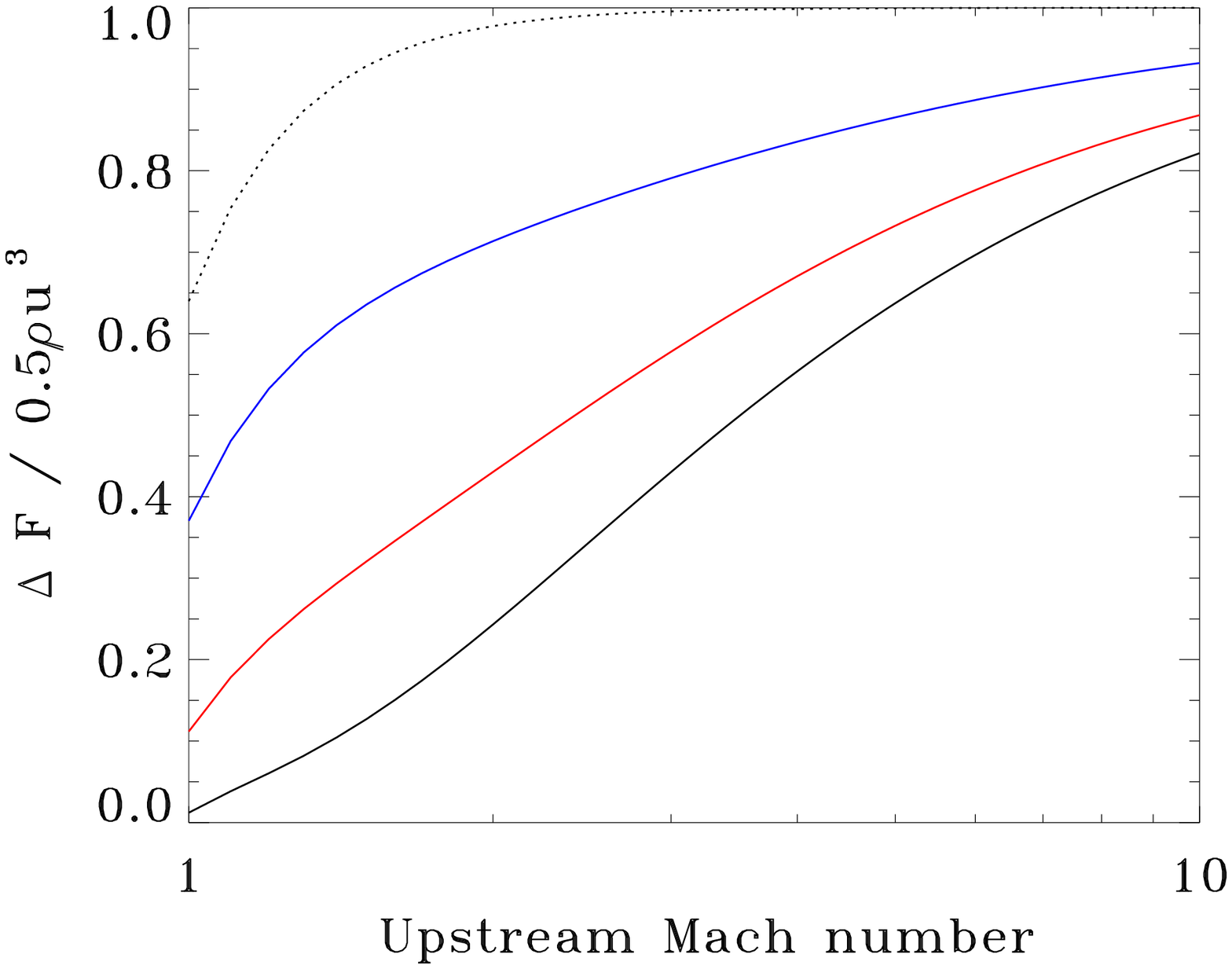}
  \caption{Evolution of  the compression ratio,  $r$, and of the  amount of
  kinetic energy radiated away, $X$, as a function of the upstream Mach number for a
  barotropic  material,  with  density  ratio $r_1=0.1$
  (black), $1$ (red), and $10$ (blue). The dotted lines represents the
  evolution for  a supercritical shock with a  optically thin upstream
  material (c.f. Fig. \ref{PX_fM}).}
\label{PX_fM_baro}
\end{figure}

Figure \ref{PX_fM_baro} illustrates the  evolution of the adimensional quantities
$r$ and $X$ as a function of the upstream Mach number  for a barotropic material,
for density  ratio $r_1=0.1$, $1$,  and $10$, which represent transition values between the optically thin (isothermal) and thick regimes (adiabatic). For comparison, we also plot
again the evolution of $r$ and $X$ for a supercritical shock with an optically thin upstream material.
The compression  ratio is not bound by a limiting value as
the upstream Mach number increases, as for the case of a supercritical shock with an optically thin upstream material, but in contrast to the case of
an opaque material. The compression ratio decreases as $r_1$ increases
(note  that for  $r_1=10$, the  material should  be  totally optically
thick in the context of star formation). The   shock  is  never   found   to  be  supercritical ($X\sim1$), even though  almost all  the incident kinetic  energy is
assumed to be  radiated away at high upstream  Mach number. This clearly shows
that  a barotropic  EOS  approximation  cannot treat properly 
radiative  shocks.  Compared to
the  case of  a supercritical  shock with  an optically  thin upstream which is relevant for the accretion shock on the first Larson core in the transition regions between the isothermal and adiabatic regime, the  compression ratio  and the
amount of energy radiated  away are underestimated.

The specific entropy jump at the shock is expressed as 
\begin{equation}
\Delta s=s_2-s_1=C_\mathrm{v} \left[ \ln\left(\frac{P_2}{\rho_2^\gamma}\right) - \ln\left(\frac{P_1}{\rho_1^\gamma}\right) \right],
\end{equation}
where $s$  is the specific entropy and $C_\mathrm{v} =
k_\mathrm{B}/(\mu  m_\mathrm{H}(\gamma-1))$. The  entropy  jump for  a
barotropic  fluid  is  thus  proportional  to  $\Pi/r^\gamma$.  Figure
\ref{deltas_baro}  shows   the  evolution  of   log$(\Pi/r^\gamma)$  for  a
barotropic fluid  (with $r_1=0.1$, 1, and  10) and for a  shock with an
optically thin  upstream material.  The entropy jump  at the  shock is
 overestimated  with a  barotropic law. This would lead to
an incorrect initial entropy level and profile for new born protostars. The entropy is indeed a key quantity for 
pre-main sequence evolutionary models as it entirely determines the mass-radius relationship of an adiabatic object, like the Larson cores.\\

\begin{figure}[tb]
  \centering
  \includegraphics[width=7cm,height=5.5cm]{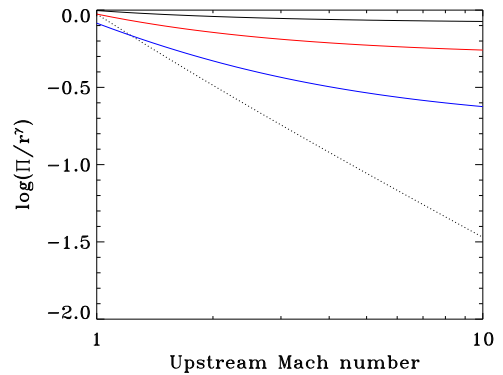}
  \caption{Evolution   of   the  entropy   jump,   $\Delta  s   \propto
  \Pi/r^\gamma $, as  a function of the the upstream  Mach number for a
  barotropic  material,  with  density  ratio $r_1=0.1$
  (black), $1$ (red), and $10$ (blue). The dotted line represents the
  evolution for  a supercritical shock with an  optically thin upstream
  material.}
\label{deltas_baro}
\end{figure}

In this section, we have shown  that the downstream quantities of  radiative shocks
strongly depend on the nature of the shock and on the model used
to describe radiation  transport.  In  the  following sections,  we  study  in detail the case of
a particular  radiative shock:  the
accretion shock on the first Larson core,  in  the context  of  star formation.


\section{1D spherical numerical calculations - Method.\label{num_cal}}

\subsection{Introduction and previous work}
In this subsection, we introduce  our numerical  method and  basics concepts,
for a good understanding of the following part of this work.

The  two  companion papers  \cite{Masunaga_Miyama_Inutsuka_I_1998ApJ},
and \cite{Masunaga_Inutsuka_2000} present a very extensive study
of 1D protostellar  collapse.  The  first paper focusses  on the formation and the properties of the first core,
using  a frequency dependent  RHD model,
while the second paper addresses the formation of the second core  These authors
use a 1D spherical code  in the Lagrangean frame.  Moment equations of
radiation    in   the   comoving    frame   are    solved   following
\cite{Stone_MM_92}.   \cite{Masunaga_Miyama_Inutsuka_I_1998ApJ} use a variable Eddington  tensor  factor   (VETF)  that   retains  a frequency 
dependence, whereas grey opacities are used to calculate the coupling between matter
and radiation and the work of the radiative pressure.

In  \cite{Masunaga_Miyama_Inutsuka_I_1998ApJ}, the cores initially  have 
uniform density  and temperature,  and a radius  adjusted so  that the
cloud should  be initially  slightly  more   massive  than   the  Jeans
mass. Initial masses range from $0.1$ M$_\odot$ to $3$ M$_\odot$. They
find that  whatever the initial cloud  mass, the  first core radius
$R_\mathrm{fc}$ is  almost constant, with $R_\mathrm{fc}\sim  5$ AU, and the  first core
mass is $M_\mathrm{fc}\sim 0.05$ M$_\odot$. In their study, $R_\mathrm{fc}$ is defined
as the point where the gas pressure is balanced by the ram pressure of
the infalling envelope. \\

To compare our numerical results with analytical theories, it is useful to define some measurable quantities. 
A  useful one is the mass accretion rate $\dot{M}$. For a 1D spherical model, the mass accretion rate is simply 
$\dot{M}=4\pi r^2 \rho u$. 
In the theory, the accretion rate is generally defined as \citep[e.g.][]{SST}
\begin{equation}
\dot{M}=\xi \frac{c^3_{s0}}{G},
\label{alpha_sst}
\end{equation}
where  $\xi$  is  a  dimensionless  coefficient  and  $c_{s0}$  the
isothermal  sound  speed.  $\xi$   is  estimated  assuming  that  a
non-magnetic,  non-rotating  cloud,  whose initial  state  corresponds
approximately to a balance between thermal  support and self-gravity, has comparable
free-fall   time, $t_\mathrm{ff}\sim  R_c^{3/2}/(GM_c)^{1/2}$,   and
sound-crossing  time, $t_\mathrm{sc}\sim  R_c/c_{s0}$. The mass
accretion  rate is thus given by  $\dot{M}\sim
M_c/t_\mathrm{ff}$.   \cite{Shu_1977} obtained  $\xi  =0.975$ for  the
expansion-wave   solution,   whereas  the
dynamic Larson-Penston solution yields $\xi   \sim  50$ \citep{Larson_1969,Penston_1969}.

Another quantity directly comparable to the theory is the  accretion  luminosity, usually
 expressed as
\begin{equation}
L^\mathrm{acc}_\mathrm{fc}=\frac{G M_\mathrm{fc}\dot{M}}{R_\mathrm{fc}},
\label{Lacc}
\end{equation}
where  $M_\mathrm{fc}$ corresponds to the
mass of  the accreting core. This accretion luminosity thus corresponds to the case where all the infalling kinetic energy is radiated away, i.e. to a supercritical radiative shock.
The accretion luminosity can be directly measured in numerical calculations, from the mass accretion rate, 
and compared with the intrinsic effective luminosity emerging from the first core.

\subsection{Numerical method}

We  use  a  1D  full  Lagrangean  version of  the  code  developed  by
Chi\`{e}ze    and   Audit    \citep[see
][]{Audit_et_al_2002}, that integrates the equations of grey radiation
hydrodynamics  under   three  different  assumptions   (in  order of increasing
complexity order):  a barotropic EOS,
the flux  limited diffusion approximation (FLD)  and the M1 model.   The RHD
equations are integrated in  their non-conservative forms using finite
volumes and an artificial viscosity scheme in tensorial form.  The RHD
equations are integrated with an implicit scheme in time, using a
standard Raphson-Newton iterative method.

\subsection{General assumptions for the radiation field}

The first  approximation in our calculations is to consider a grey radiation  field, i.e.
only  one  group   of  photons,  integrated   over  all
frequencies. 
We also  neglect scattering and
assume Local Thermodynamical Equilibrium (LTE) everywhere.
The second  assumption concerns  the coupling between radiation and  hydrodynamics, for which we write  the RHD  equations  in the comoving frame.

\subsection{Models for the radiative transfer}

According to  previous studies using accurate model  for the radiation
field  \citep[e.g.][]{Masunaga_Miyama_Inutsuka_I_1998ApJ}, it  is well
established that the molecular  gas follows two thermal regimes during
the first collapse. At low density,  the gas is able to radiate freely
and to couple with the dust.  This is the isothermal regime, where the Jeans
mass decreases  with density.  Then, when the gas  becomes denser,
typically $\rho  > \rho_\mathrm{ad}\sim 1\times10^{-13}$  g cm$^{-3}$,
the radiation  is trapped  within the  gas that begins  to heat up
adiabatically.  The Jeans mass  then increases with density until
the gas becomes hot enough to dissociate H$_2$, leading to the onset of the second
collapse.

\subsubsection{The barotropic equation of state approximation \label{EOS_barotrop}}

As mentioned earlier, the easiest way to describe the thermal evolution of the gas  without solving
the radiative transfer equation is to adopt a barotropic EOS that
reproduces both the isothermal  and adiabatic limiting regimes as a  function of the
density. We use the following barotropic EOS:
\begin{equation}
\frac{P}{\rho} = c_\mathrm{s}^2 = c_{s0}^2\left[1+\left(\frac{\rho}{\rho_\mathrm{ad}}\right)^{2/3}\right],
\label{baro}
\end{equation}
where  $\rho_\mathrm{ad}$ is  the critical  density at  which  the gas
becomes adiabatic.  This critical density is obtained by more accurate
calculations and depends on the  opacities, the composition and the
geometry  of   the  molecular  gas.   At  low   densities,  $\rho  \ll
\rho_\mathrm{ad}$, $c_\mathrm{s}  \sim c_{s0}=0.19 $  km s$^{-1}$ (for $T\sim 10$ K), the
molecular gas is  able to radiate freely by  thermally coupling to the
dust  and remains  isothermal  at 10  K.   At high  densities, $\rho  >
\rho_\mathrm{ad}$,  we  assume  that  the  cooling  due  to  radiative
transfer  is  trapped  by  the  dust  opacity.   Therefore,  $P\propto
\rho^{5/3}$,  which corresponds  to an  adiabatic, monoatomic  gas with
adiabatic exponent  $\gamma = 5/3$. Molecular hydrogen  behaves like a
monoatomic gas  until the temperature reaches  several hundred Kelvin since
the rotational degrees of freedom  are not excited at
lower  temperatures
\citep{Whitworth_Clarke_1997,Masunaga_Inutsuka_2000}.

\subsubsection{Moment Models}

In  RHD  calculations,  the  radiative  transfer  equation  should  be
formally integrated over 6 dimensions  at each time step. This process
is  too  computationally   demanding  for  multidimensional  numerical
analysis.  Using angular  moments of the  transfer equation is thus very useful, by allowing  a large
reduction of the computational  cost. However, each evolution equation
of a  moment of the transfer  equation involves the  next higher order
moment of  the intensity. Consequently,  as for the kinetic  theory of
gases, the  system must be  closed by using  an {\it ad  hoc} relation
that  gives  the highest  moment  as a  function  of  the lower  order
moments.  For  radiation  transport,  the closure  theory  is  usually
limited to the  two first moments of the  transfer equation. A closure
relation  for the  system  is needed  and  this relation  is of  prime
importance. Many  possible choices for the closure  relation exist. In
the following,  we present two models  based on more  or less accurate
closure  relations,  that  we  use  in this  work:  the  flux  limited
diffusion (FLD) approximation and the M1 model.

\subsubsection{The Flux Limited Diffusion approximation\label{sec_FLD}}

The diffusion  approximation is the most widely used  moment model of
radiation  transport.  The diffusion  limit is  valid when  the photon
mean  free path  is small  compared with  other length  scales  in the
system. On  the contrary, the  approximation is no longer  accurate in
the transport regime.  In the diffusion limit, photons diffuse through
the  material in  a  random walk.  Readers  can find  an  accurate
derivation of the diffusion limit in \cite{Mihalas_book}, \S 80.

In  the   diffusion  limit,  
the radiative energy $E_\mathrm{r}$ and
radiative flux $\textbf{F}_\mathrm{r}$ are simply related by
\begin{equation}
\textbf{F}_\mathrm{r} = -\frac{c}{3\kappa_\mathrm{R}}\nabla E_\mathrm{r},
\label{diff}
\end{equation}
where $\kappa_\mathrm{R}$  is the  {\it Rosseland mean  opacity}.  The
radiative flux  is expressed directly  as a function of  the radiative
energy  and  is proportional  and  collinear  to  the radiative  energy
gradient.  Equation (\ref{diff}) has  no upper limit, but
for  optically thin  flows,  the effective  propagation  speed of  the
radiation   must   be   limited  to   $c$   ($\textbf{F}_\mathrm{r}\le
cE_\mathrm{r}$). We  thus have to  limit the propagation speed  of the
radiation by means  of a flux limiter.  Equation  (\ref{diff}) is then
expressed as
\begin{equation}
\textbf{F}_\mathrm{r} = -\frac{c\lambda}{\kappa_\mathrm{R}}\nabla E_\mathrm{r},
\label{fld_closure}
\end{equation}
where  $\lambda  =  \lambda  (R)$  is  the  flux  limiter. 

In this study, we retain the flux limiter 
derived by \cite{Minerbo_1978JQSRT}\begin{equation}
\lambda=\left\{\begin{array}{ccc}
2/(3+\sqrt{9 + 12 R^ 2}) &  \mathrm{if } &    0\le R \le 3/2,\\
(1+ R + \sqrt{1 + 2 R})^{-1} & \mathrm{if } &  3/2 <  R \le \infty,
\end{array}\right.
\end{equation}
with $R=|\nabla  E_\mathrm{r} |/(\kappa_\mathrm{R} E_\mathrm{r})$. The  flux  limiter  has  the property  that  $\lambda
\rightarrow 1/3$  in optically thick regions  and $\lambda \rightarrow
1/R$  in optically  thin regions.   

In the FLD approximation, an unique diffusion-type equation on the radiative energy is thus obtained
\begin{equation}
\label{nrjrad} 
\frac{\partial E_\mathrm{r}}{\partial t} - \nabla \cdot\left(\frac{c\lambda}{\kappa_\mathrm{R}} \nabla E_\mathrm{r}\right) = 
\kappa_\mathrm{P}(4\pi B -cE_\mathrm{r}).
\end{equation}

\subsubsection{M1 model}

In the M1 model, the radiation transport is described by the first two moments of the radiative transfer  equation (radiative  energy and
flux). We use a closure relation introduced by \cite{Dubroca_1999}. The M1 method is  used in the RHD code {\ttfamily HERACLES} \rm \citep{Gonzalez_2007}.
Based  on a  minimum  entropy  principle, this  method  is able  to
account  for large  anisotropy of  the radiation  as well  as  for the
correct diffusion limit.  The main  advantage of the M1 system is that the
underlying photon  distribution function is  not isotropic, but  has a
preferential direction of propagation. The M1 model also allows to explicitly  get rid of the {\it ad-hoc} limitation of the flux in the transport regime.

Let us consider the evolution equations of the zeroth and first moments
of the specific intensity in the laboratory frame

\begin{equation} 
\left\{
\begin{array}{llll}
\partial_t E_\mathrm{r} + \nabla\cdot\textbf{F}_\mathrm{r} & = &\kappa_\mathrm{P}(4\pi B -c E_\mathrm{r})\\
\partial_t \textbf{F}_\mathrm{r} + c^2\nabla\cdot\mathbb{P}_\mathrm{r}  &=& -\kappa_\mathrm{R} c\textbf{F}_\mathrm{r}
\end{array}
\right. ,
\label{M1}
\end{equation}
where  $\kappa_\mathrm{P}$ is the {
\it Planck  mean opacity}.  Note that we use the Rosseland mean in the first moment equation in order to yield the  diffusion limit. 
As a closure relation, the     radiative      pressure     is     often      expressed     as
$\mathbb{P}_\mathrm{r}=\mathbb{D}E_\mathrm{r}$,  where  $\mathbb{D}$ is
the  Eddington tensor. Assuming  that the  direction of  the radiative
flux  is an  axis of  symmetry of  the local  specific  intensity, the
Eddington tensor is given by \citep{Levermore_1984JQSRT}
\begin{equation}
\mathbb{D}=\frac{1-\chi}{2}\mathbb{I}+\frac{3\chi-1}{2}\textbf{n}\otimes\textbf{n},
\label{eddington}
\end{equation}
where $\chi$ is the Eddington factor, $\mathbb{I}$ the identity matrix
and  $\textbf{n}=\frac{\textbf{f}}{f}$ a unit  vector aligned  with the
radiative flux. In the M1 approximation, the Eddington factor $\chi$ is then obtained by minimizing the radiative entropy 
\citep{Dubroca_1999}. The Eddington factor is then found to be
\begin{equation}
\chi = \frac{3+4f^2}{5+2\sqrt{4-3f^2}}.
\label{m1_closure}
\end{equation}

It is then trivial to  recover the two asymptotic regimes of radiative
transfer.      When    $f     \rightarrow    0$,     $\chi=1/3$    and
$\mathbb{P}_\mathrm{r}=1/3E_\mathrm{r}\mathbb{I}$ which corresponds to
the diffusion limit with an isotropic radiative pressure. On the other
hand,            if           $f=1$,            $\chi=1$           and
$\mathbb{P}_\mathrm{r}=\textbf{n}\otimes\textbf{n}E_\mathrm{r}$, which
corresponds to the free-streaming limit. Between these two
limits, the  closure relation ensures that energy  remains positive and
that the flux is limited ($\textbf{F}_\mathrm{r}<cE_\mathrm{r}$).\\

\subsubsection{Systems of RHD equations}

\begin{itemize}
\item In the barotropic EOS approximation, the thermal behavior
of  the gas is determined by the choice of the EOS.  The energy  equation becomes  superfluous and  the Euler
equations reduce to :

\begin{equation}
\left\{
\begin{array}{llll}
\partial_t \rho + \nabla \left[\rho\textbf{u} \right] & = & 0 \\
\partial_t \rho \textbf{u} + \nabla \left[\rho \textbf{u}\otimes \textbf{u} + P \mathbb{I} \right]& =& -\rho\nabla\Phi,\\
\end{array}
\right. 
\end{equation}
where $\Phi$ the gravitational potential.\\

\item In the FLD approximation, the system of RHD equations corresponds to the Euler equations plus the equation on the radiative energy
\begin{equation}
\left\{
\begin{array}{ccccl}
\partial_t \rho & + & \nabla \left[\rho\textbf{u} \right] & = & 0 \\
\partial_t \rho \textbf{u} & + & \nabla \left[\rho \textbf{u}\otimes \textbf{u} + P \mathbb{I} \right]& =& -\rho\nabla\Phi +  \lambda\nabla E_\mathrm{r} \\
\partial_t E & + & \nabla \left[\textbf{u}\left( E + P \right)\right] &= &-\rho\textbf{u}\cdot\nabla \Phi + \lambda\nabla E_\mathrm{r} \cdot\textbf{u}  \\
 & & & & - \kappa_\mathrm{P}(4\pi B -cE_\mathrm{r})\\
\partial_t E_\mathrm{r} & + & \nabla \left[\textbf{u}E_\mathrm{r}\right]
 &=&   -  \mathbb{P}_\mathrm{r}:\nabla\textbf{u} + \nabla \cdot\left(\frac{c\lambda}{ \kappa_\mathrm{R}} \nabla E_\mathrm{r}\right)\\
 & & & & + \kappa_\mathrm{P}(4\pi B -cE_\mathrm{r}).
\end{array}
\right.
\end{equation}
This system is closed by the perfect gas relation and the simplest FLD approximation for the radiative pressure, $\mathbb{P}_\mathrm{r}=\lambda E_\mathrm{r}$.
\\

\item In the M1 model, we consider the equations governing the evolution of an inviscid, radiating fluid 
\begin{equation}
 \left\{
\begin{array}{ccccl}
\partial_t \rho & + & \nabla \left[\rho\textbf{u} \right] & = & 0 \\
\partial_t \rho \textbf{u} & + & \nabla \left[\rho \textbf{u}\otimes \textbf{u} + P \mathbb{I} \right]& =& -\rho\nabla\Phi + \kappa_
\mathrm{R}\textbf{F}_\mathrm{r}/c \\
\partial_t E & + & \nabla \left[\textbf{u}\left( E + P \right)\right] &= &-\rho\textbf{u}\cdot\nabla \Phi + \kappa_\mathrm{R}\textbf{F}_\mathrm{r}/c\cdot\textbf{u}\\
 & & & &  - \kappa_\mathrm{P}(4\pi B -cE_\mathrm{r})\\
\partial_t E_\mathrm{r} & + & \nabla \left[\textbf{u}E_\mathrm{r}\right]  + \nabla\cdot\textbf{F}_\mathrm{r}& = &  -  \mathbb{P}_\mathrm{r}:\nabla\textbf{u} + \kappa_\mathrm{P}(4\pi B -c E_\mathrm{r})\\
\partial_t \textbf{F}_\mathrm{r} & + & \nabla \left[\textbf{u}\textbf{F}_\mathrm{r}\right]  + c^2\nabla\cdot\mathbb{P}_\mathrm{r}& = & - 
\left(\textbf{F}_\mathrm{r}\cdot\nabla\right)\textbf{u}  -\kappa_\mathrm{R} c\textbf{F}_\mathrm{r}.
\end{array}
\right.
\end{equation}
where $\rho$ is the material density, $\textbf{u}$ is the velocity, $P$ the isotropic thermal pressure, $\Phi$ the gravitational potential and $E$ the fluid total energy $E=\rho\epsilon +1/2\rho u^2$. This system is closed by the perfect gas relation and the M1 relation  (\ref{m1_closure}).

\end{itemize}

\subsection{Initial and boundary conditions}
 We         use          the         same         model         as
\cite{Masunaga_Miyama_Inutsuka_I_1998ApJ},  i.e.   a  uniform  density
sphere   of   mass   $M_0=1$   M$_\odot$,   temperature   $T_0=10$   K
($c_{s0}=0.187$ km  s$^{-1}$) and radius $R_0=10^4$  AU.  This initial
setup  corresponds to  a ratio  $\alpha$ of  thermal  to gravitational
energies of  $\alpha=5R_0 k_\mathrm{B}T_0/(2 G  M_0 \mu m_\mathrm{H})
\sim 0.97$ and to a free-fall time t$_\mathrm{ff} \sim 1.77 \times 10^
5$  yr.  Boundary conditions  are very  simple: for  hydrodynamics, we
impose  a constant  thermal pressure  equal to  the  initial pressure
(other quantities are  free) and for the radiation  field, we impose a
vanishing gradient  on radiative temperature.   Calculations have been
performed using a Lagrangean grid containing 4500 cells.

\subsection{The opacities}

For  the  M1 model  and  the  FLD approximation,  we  use  the set  of
opacities  given by  \cite{Semenov_et_al_2003A&A} for  low temperature
($<1000$ K)  and \cite{Ferguson_05} for high  temperature ($>1000$ K),
that we compute as a function of the gas temperature and density.  
In Fig. \ref{table_opr}, the table for the Rosseland opacity
is plotted as a function of temperature and density. 
  In
\cite{Semenov_et_al_2003A&A},   the  dependence  of   the  evaporation
temperatures of ice, silicates and  iron on gas density are taken into
account. In this work,  we use spherical composite aggregate particles
for the grain structure and topology  and a normal iron content in the
silicates, Fe/(Fe + Mg)=0.3.

\begin{figure}
  \centering \includegraphics[width=8cm,height=6cm]{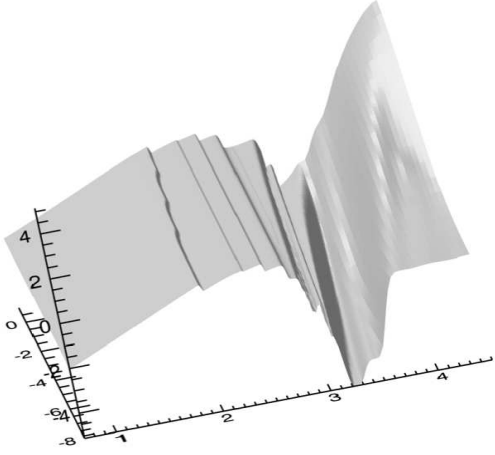}
  \caption{Rosseland     opacity     made      of     a     mix     of
  \cite{Semenov_et_al_2003A&A}   model    at   low   temperature   and
  \cite{Ferguson_05} model  at high temperature.  The Rosseland mean opacity is
  plotted    as   a   function    of   temperature    ($x$-axis)   and
  $R=\rho/(T^3_6)$,  with  $T_6=T/10^6$,  ($y$-axis)  using  logarithm
  scales. }
\label{table_opr}
\end{figure}

\section{1D numerical calculations of the first collapse\label{results}}

As  we mentioned  before, all  calculations presented  here  have been
performed  with 4500  cells,  distributed according  to a  logarithmic
scale  in  mass  in  our   initial  setup.   These  mass  and  spatial
resolutions are  sufficient to resolve quantitatively the  accretion shock energy budget (see appendix \ref{appendixA}).

Calculations with the barotropic EOS have been performed
using  a critical density  of $\rho_\mathrm{ad}=3.7\times  10^{-13}$ g
cm$^{-3}$ .  We derived the  value of $\rho_\mathrm{ad}$  from
the intersection of extrapolated lines  of the isotherm and the adiabat in
the log($\rho$)-log($T$)  plane, obtained with  calculations using the
M1 model (see Fig. \ref{plot_rhoad}).

\begin{figure}[tb]
  \centering
  \includegraphics[width=7cm,height=5.5cm]{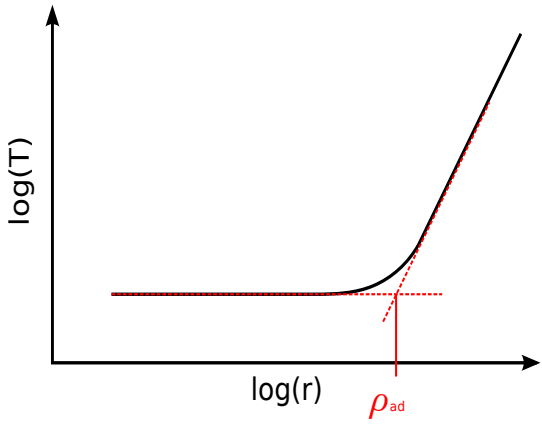}
    \caption{Sketch of the estimate of $\rho_\mathrm{ad}$ used in the barotropic EOS, 
    according to the density-temperature distribution obtained with the  M1 model.}
\label{plot_rhoad}
\end{figure}

\subsection{Results during the first collapse. Formation of the first core.}

Table \ref{M1_DIFF} gives results of barotropic EOS (Baro), M1, and FLD
calculations,   when    the   central   density    reaches   $\rho_c   =
1\times10^{-10}$  g  cm$^{-3}$.    Note  that  for  all  calculations,
the dynamical  times  are very  close  to  $\sim  0.189$ Myr  $\sim  1.07$
t$_\mathrm{ff}$.  We  define the  first core radius  as the  radius at
which  the infall  velocity  is  the largest  (i.e.,  the position of  the
accretion shock).  The accretion luminosity is estimated  at the first
core border, according to  equation (\ref{Lacc}). The mass
accretion  rate  $\dot{M}$ and  the  accretion  parameter $\xi$  are evaluated at
$R_\mathrm{fc}$.   Table 1 also displays the  central
temperature  $T_\mathrm{c}$, the central specific entropy  $s_\mathrm{c}$, and the first core temperature at the border $T_\mathrm{fc}$.

As seen from table \ref{M1_DIFF}, the first core radius and mass agree well  with \cite{Masunaga_Miyama_Inutsuka_I_1998ApJ} results. The
results from  the M1 or  FLD calculations are very similar. The
central temperature with the FLD is  slightly higher (by $\sim 2\%$) than the
one in  the M1  calculations, which confirms the
fact  that the  diffusion approximation tends  to slightly overestimate the cooling. The central entropy obtained with the M1 and FLD
models is lower than the one obtained with the barotropic EOS, and  this difference increases with time. Indeed,
taking into account radiative  transfer allows the energy to be radiated  away during the collapse, a process which is not properly 
handled  with a barotropic  EOS. This
leads to a lower  entropy level of the first core with  the FLD or the
M1 models. The temperature at the border of the first core is higher
in the  M1 and FLD cases,  since the photons escaping  the accretion shock
heat up the  infalling  material.  On the other hand, the  mass
accretion  rate,  the mass,  the  radius  and  the  accretion
parameter $\xi$ of the  first core are in good agreement  between  the three models.
Note that the values of $\xi$ obtained in our calculations are closer to the Larson-Penston solution than to Shu's
expansion wave solution.

\begin{table*}[tb]
\center
\begin{tabular}{ccccccccc}
\hline
\hline
Model  & $R_\mathrm{fc}$ & $M_\mathrm{fc}$ &$\dot{M}$             & $L_\mathrm{acc}$ & $T_\mathrm{c}$ & $T_\mathrm{fc}$ & $s_\mathrm{c}$  &$\xi$\\
             &                        (AU) &   (M$_\odot$)       & (M$_{\odot}/$yr) &   (L$_\odot$)         &  (K) & (K) & (erg K$^{-1}$ g$^{-1}$) & \\
\hline
\small 
Baro   &  6.8  &  $2.35 \times 10^{-2}$  &  $4.2 \times 10^{-5}$  &  0.021  &  551  &  15  & $2.03 \times 10^9$&27\\ 
 FLD   &  8.1  &  $2.33 \times 10^{-2}$  &  $4.1 \times 10^{-5}$  &  0.019  &  411  &  58  & $2.02 \times 10^9$& 26\\
   M1   &  7.8  &  $2.33 \times 10^{-2}$  &  $4.1 \times 10^{-5}$  &  0.018  &  419  &  57  & $2.02 \times 10^9$& 26\\ 
      \hline
\end{tabular} 
\caption{Summary of first core properties for $\rho_c=1\times 10 ^{-10}$ g cm$^{-3}$.}
\label{M1_DIFF}
\end{table*}

\begin{figure*}[tb]
  \centering
  \includegraphics[width=7cm,height=5.5cm]{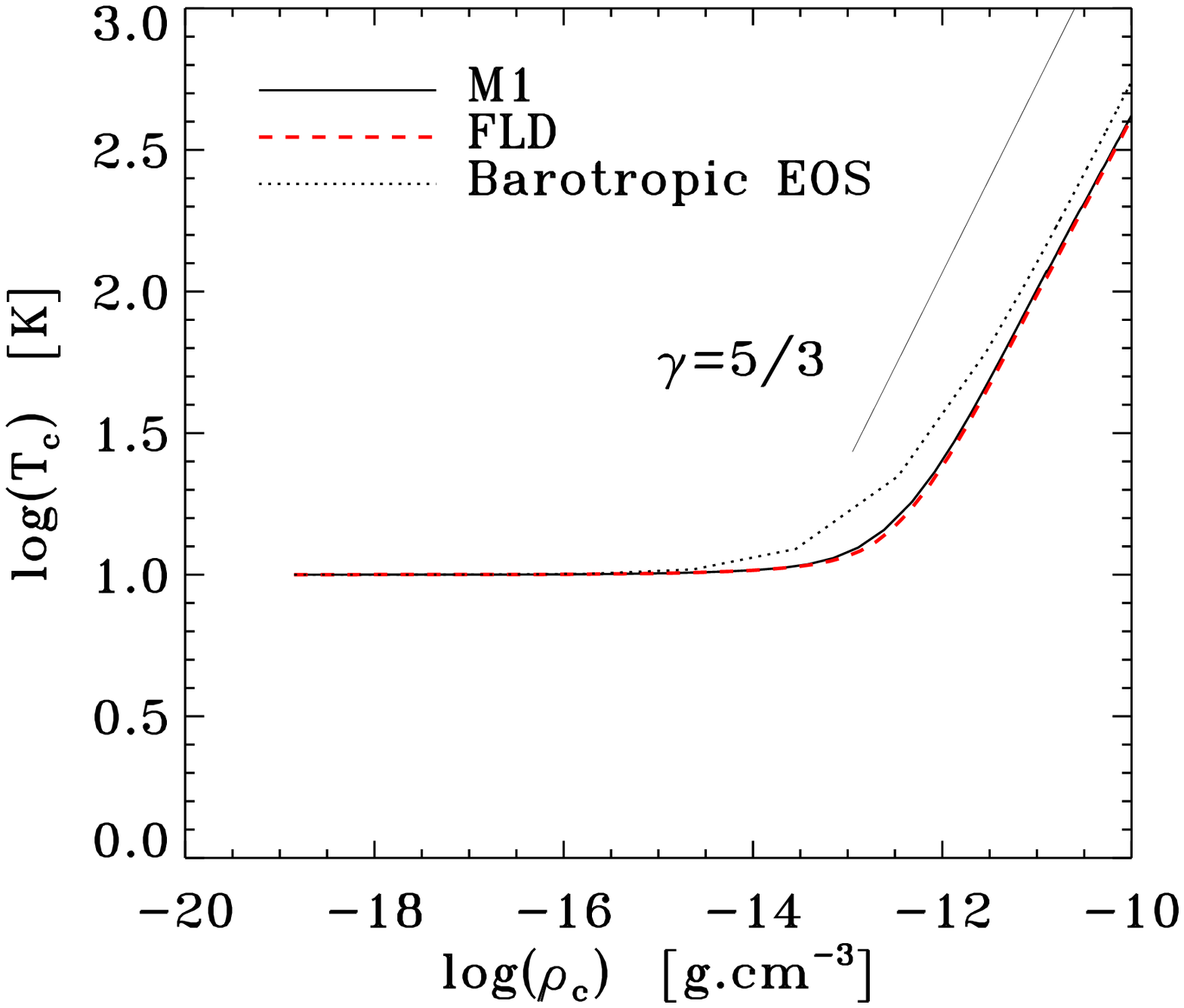}
  \includegraphics[width=7cm,height=5.5cm]{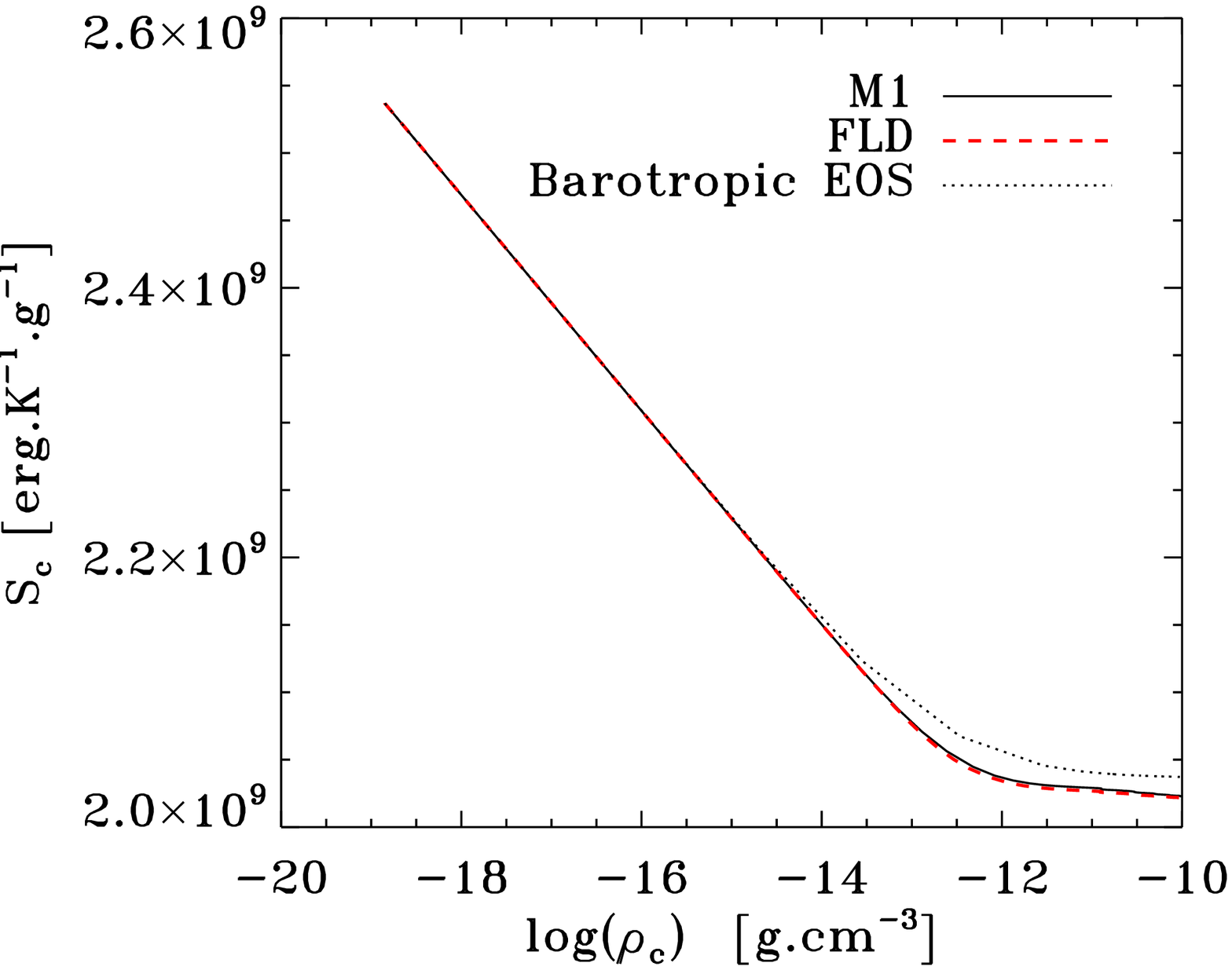}
  \caption{Evolution of the central temperature ( {\it left}) and the central entropy  ({\it right}) as 
    a function of the central density during the first collapse for  the M1 (solid line),  FLD (dashed red line) and  barotropic (dashed-dotted line).}
\label{plot_centre}
\end{figure*}

Figure   \ref{plot_centre}  shows   the  evolution   of   the  central
temperature and the central entropy 
as a  function of the
central  density.   From  Fig.  \ref{plot_centre}(a),  we
notice    the   perfect   bimodal thermal   behavior    of    the   gas, from isothermal to adiabatic. The  critical density  at which  the gas
begins to  heat in both M1  and FLD calculations is  the same.  The
difference with the barotropic case stems from our choice
of the barotropic  EOS and critical  density. For FLD
and M1, we  find a slope of $\gamma_\mathrm{eff} - 1  \sim 0.61 < 2/3$
at   high   density. This  means that the first  core is not completely adiabatic, but
experiences some heat loss, yielding a significant cooling.

In Fig.  \ref{plot_centre}(b), the central  entropy is plotted  as a
function  of the  central density.   At high  density, we  recover the
adiabatic  regime, and the M1 and  FLD  calculations settle  at the  same
entropy level in the center, that is lower than the one reached by the
barotropic model. In the barotropic case, the entropy is determined by the
EOS. The value of the ``adiabat" at which the gas settles is
\begin{equation}
s  \propto \ln\left(\frac{P}{\rho^\gamma}\right) .
\end{equation}
For the barotropic EOS, we have
\begin{equation}
  s\propto \ln\left[
\rho^{-2/3}c_{s0}^2\left(1+\left(\frac{\rho}{\rho_\mathrm{ad}}\right)^{2/3}
\right)\right]
\propto  \ln\left(  \frac{c_{s0}^2}{\rho_\mathrm{ad}^{2/3}}  +
\frac{c_{s0}^2}{\rho^{2/3}}\right).
\end{equation}
At       high       density,        $s       \rightarrow
\ln(c_{s0}^2/\rho_\mathrm{ad}^{2/3})$.  The  limiting value for  the entropy
is  then  $\sim  2.03  \times  10^9$ erg  K$^{-1}$  g$^{-1}$  for  the
barotropic EOS, which  is higher than the value  obtained with FLD and
M1. Moreover,  as the slope of the thermal profile (fig.  \ref{plot_centre}(a)) is not
exactly equal to  $\gamma=5/3$ in the M1 and FLD calculations, as mentioned above, the gas  tends to cool and  to decrease
its  entropy level at  a  rate  $\sim  10^{-3}$-$10^{-4}$  erg  K$^{-1}$  g$^{-1}$
s$^{-1}$ at the center. This  is one of  the immediate and most important consequences of  correctly taking into account
radiative  transfer in the collapse:  the  accretion shock  becomes  a  real
radiative shock and radiation is transported outward in the core. \\

\begin{figure*}[tb]
  \centering
  \includegraphics[width=7cm,height=5.5cm]{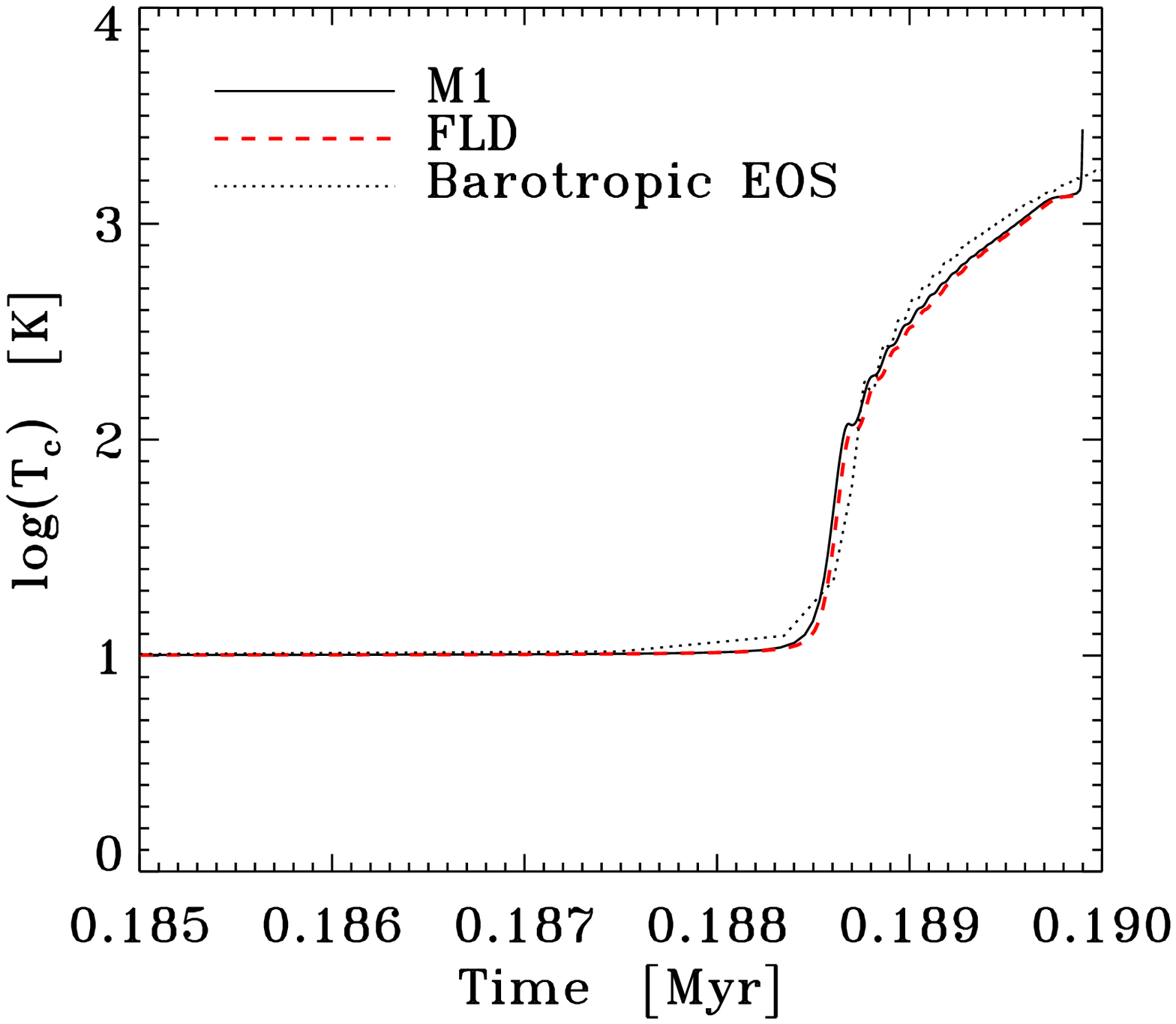}
  \includegraphics[width=7cm,height=5.5cm]{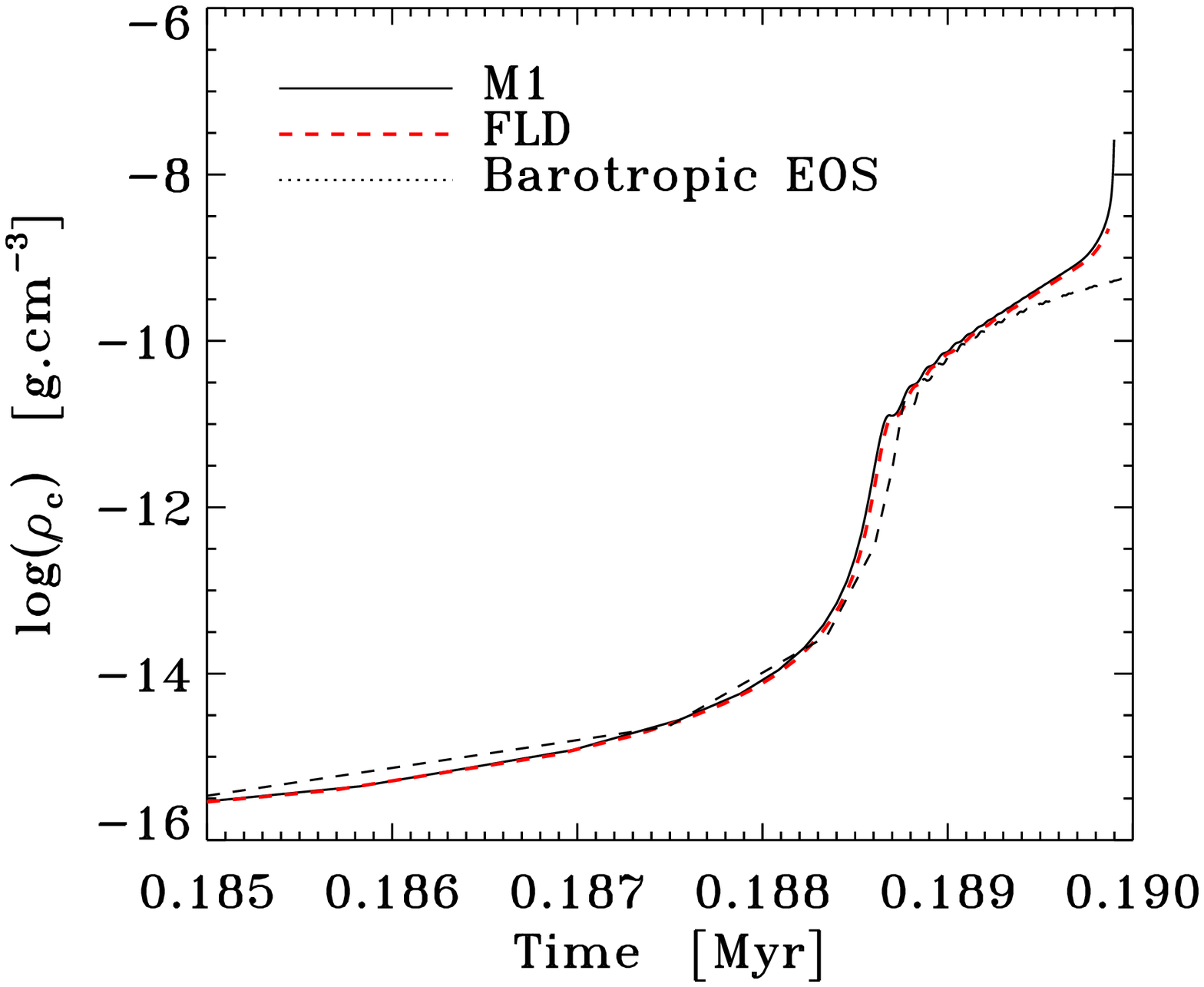}
 \includegraphics[width=7cm,height=5.5cm]{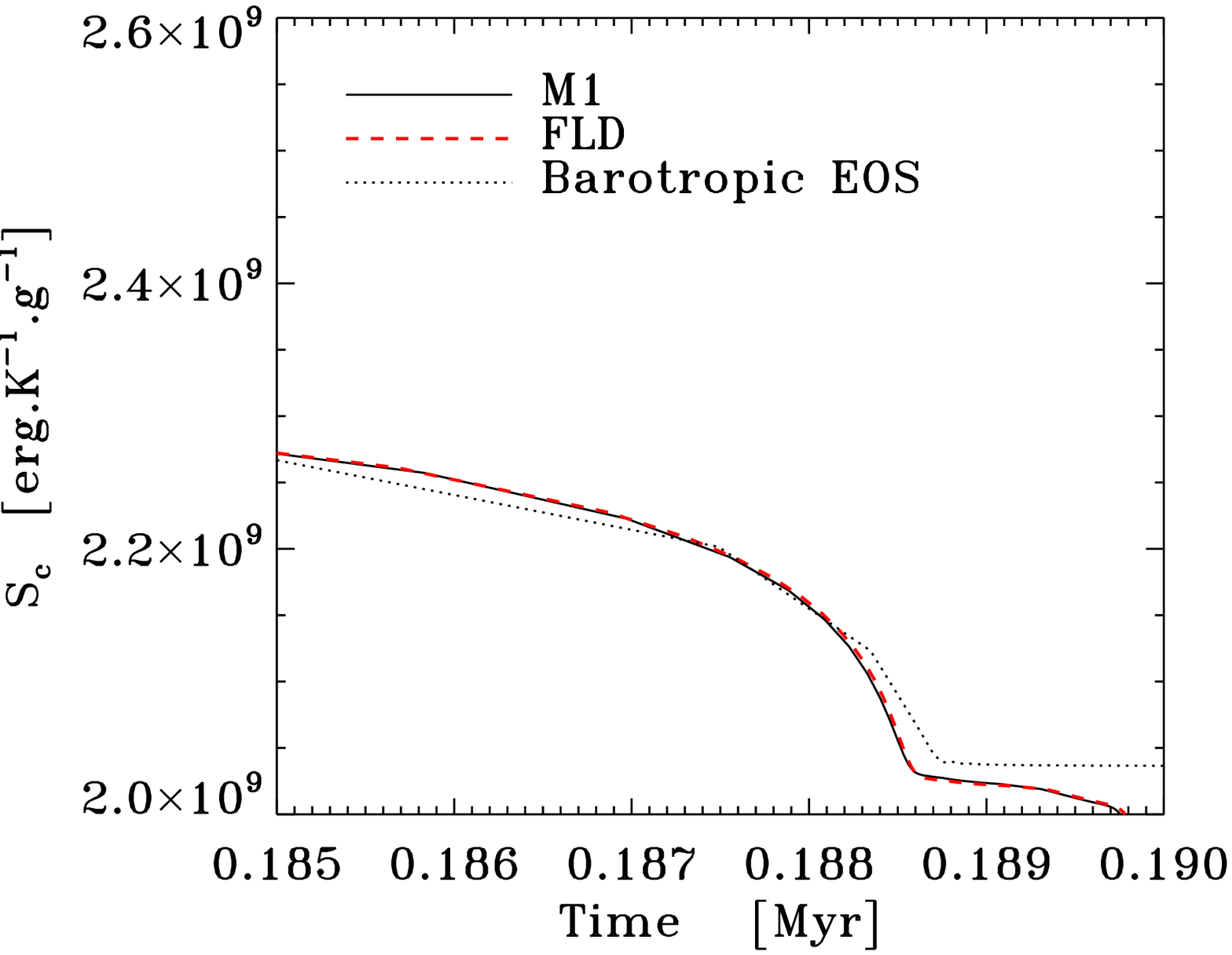}
 \includegraphics[width=7cm,height=5.5cm]{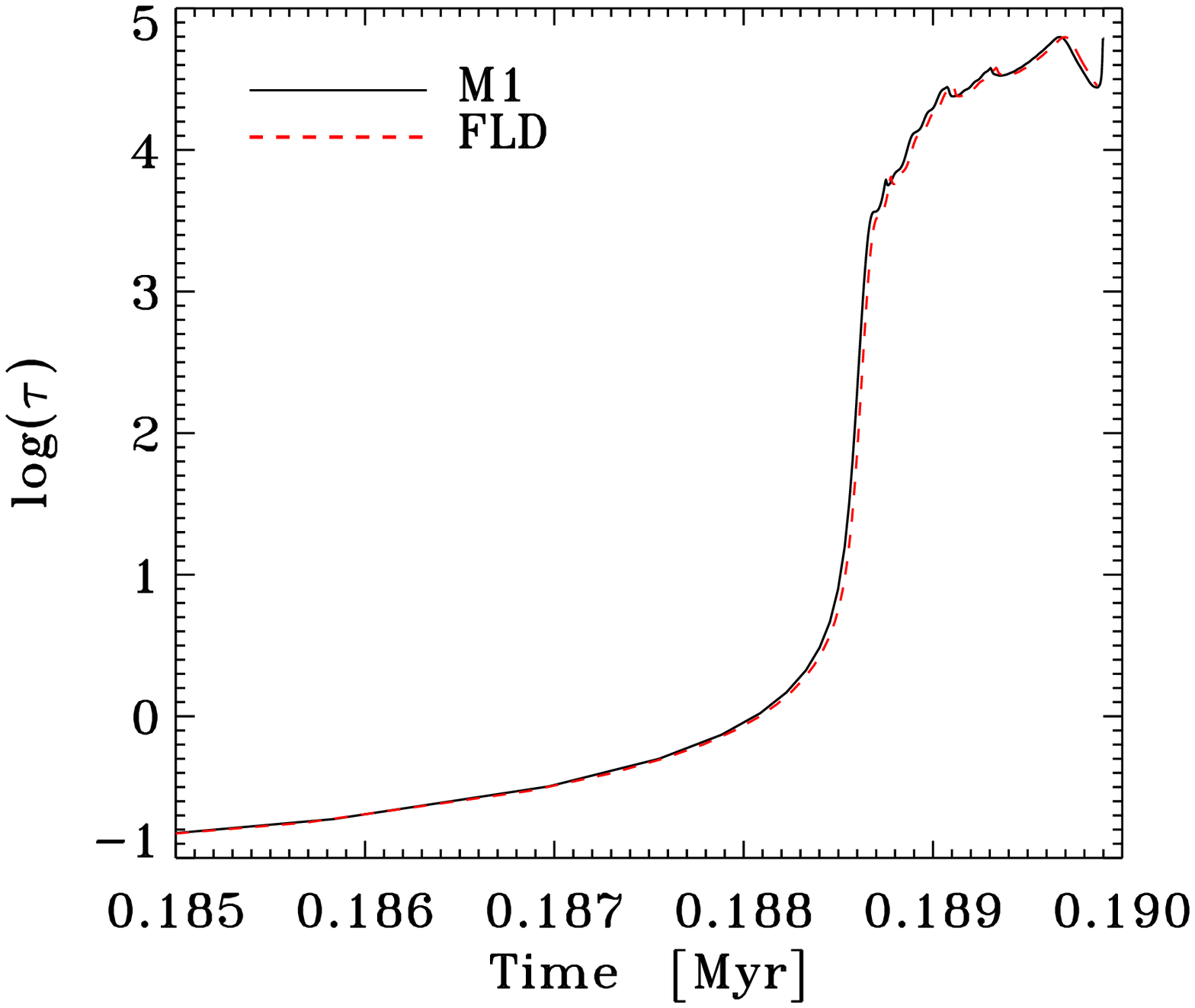}
  \caption{Evolution of the central temperature, density, entropy and optical depth with time for  the M1 (solid line),  FLD (dashed red line) and  barotropic (dashed-dotted line) models.}
\label{plot_central_time}
\end{figure*}

Figure \ref{plot_central_time} shows the central temperature, density,
entropy and optical depth evolution during the collapse. Variations of
all variables are quite similar for temperature lower than T$\sim 100$
K.  At 100  K, there is a first discontinuity  in the opacity,
due  to the destruction of icy grains.  This discontinuity moves to higher
temperature as density  increases, so that density affects to some extent the
opacity.  However,  these differences
are  really small.   From  Fig. \ref{plot_central_time}(c),  we
clearly see  that the entropy level  remains constant with  the barotropic
model, whereas the first core  keeps cooling to lower entropy
levels with the  M1 and FLD calculations.  This is an important result,  since, as mentioned previously, the
entropy level of a low mass protostar determines its mass-radius relationship, and thus its pre-main sequence subsequent
evolution.  According to the first principle of thermodynamics, the global entropy loss of the first core can be estimated as
\begin{equation}
\left(\frac{\partial L}{\partial m}\right)_t=-T\left(\frac{\partial s}{\partial t}\right)_m = -\left(\frac{\partial \epsilon}{\partial t}\right)_m-P\left(\frac{\partial v}{\partial t}\right)_m,
\end{equation}
where $\epsilon$ is the specific internal energy and $v=1/\rho$ is the specific volume.
If the first core was perfectly adiabatic, the compressional work would be entirely converted into internal energy, and the radiative loss would be zero. As shown in Fig \ref{plot_dc_1dm10}f, the luminosity increases significantly within the first core in the FLD and M1 calculations. This increase corresponds to the radiative loss, which amounts to $\sim 4\times10^{-3}$ L$_\odot$. The corresponding entropy loss of the first core is then 
\begin{equation}
\frac{\Delta s_\mathrm{fc}}{\Delta t}\sim-\frac{1}{T}\frac{\Delta L}{M_\mathrm{fc}}.
\end{equation}
At $\rho_c=1\times10^{-10}$ g cm$^{  -3}$, we get $M_\mathrm{fc}\sim 2.3\times10^{-2}$ M$_\odot$. The entropy loss is thus
\begin{equation}
\frac{\Delta s_\mathrm{fc}}{\Delta t}\sim-\frac{3.3\times10^{-1}}{T} \hspace{6pt} \mathrm{erg}\hspace{3pt}\mathrm{K}^{-1}  \mathrm{g}^{-1}.
\end{equation}
For a typical first core temperature of a few 100 K, the entropy loss rate is   $\sim 10^{-3}$ erg K$^{-1}$ g$^{-1}$ yr$^{-1}$, which is consistent with the value found at the center. The first core lasting about a few hundred years, the total entropy loss thus represents a few percents of the core's initial entropy content, and this entropy loss increases with time, as shown in Fig. 1 of \cite{Masunaga_Miyama_Inutsuka_I_1998ApJ}. Such an entropy loss cannot be handled with a barotropic EOS.

In  Fig. \ref{plot_central_time}(d),  we  see that  the
first  core quickly  becomes optically  thick once  it begins  to heat
up. The optical depth  at the center is so high  that observations are not
able to  catch the  central evolution  of the cores  at this  stage of
the evolution.

\begin{figure*}
  \centering
  \includegraphics[width=6.5cm,height=5cm]{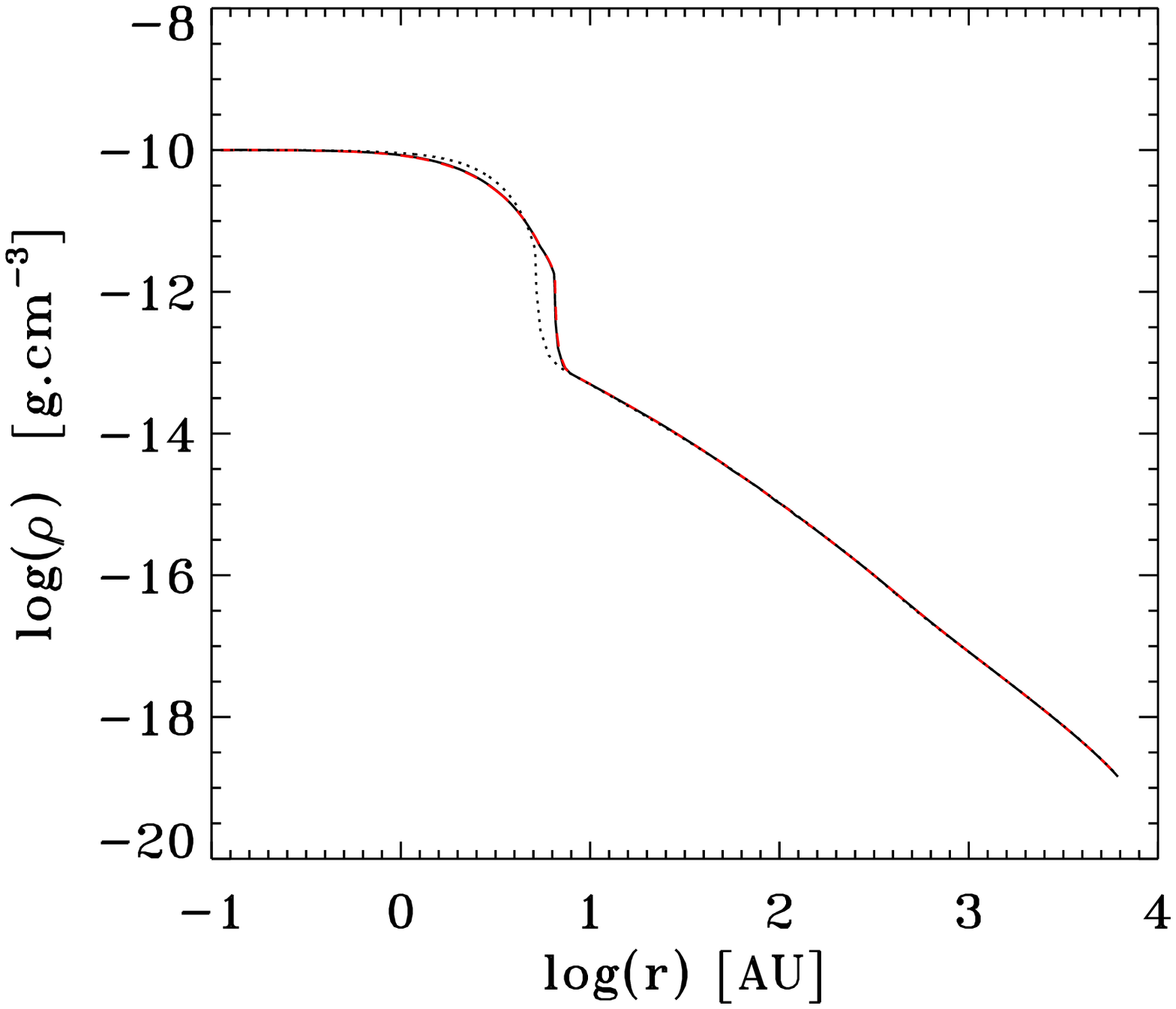}
  \includegraphics[width=6.5cm,height=5cm]{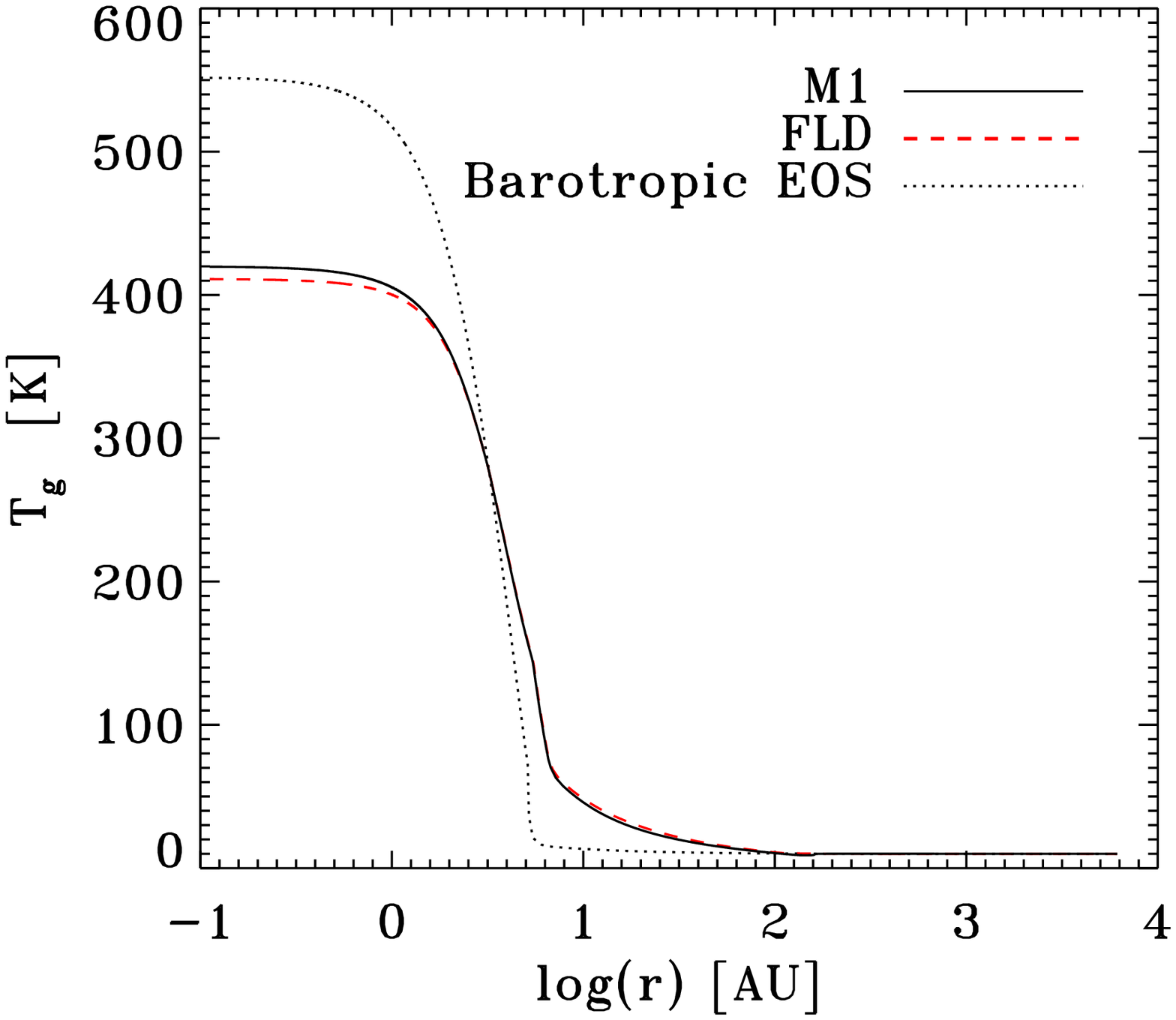}
  \includegraphics[width=6.5cm,height=5cm]{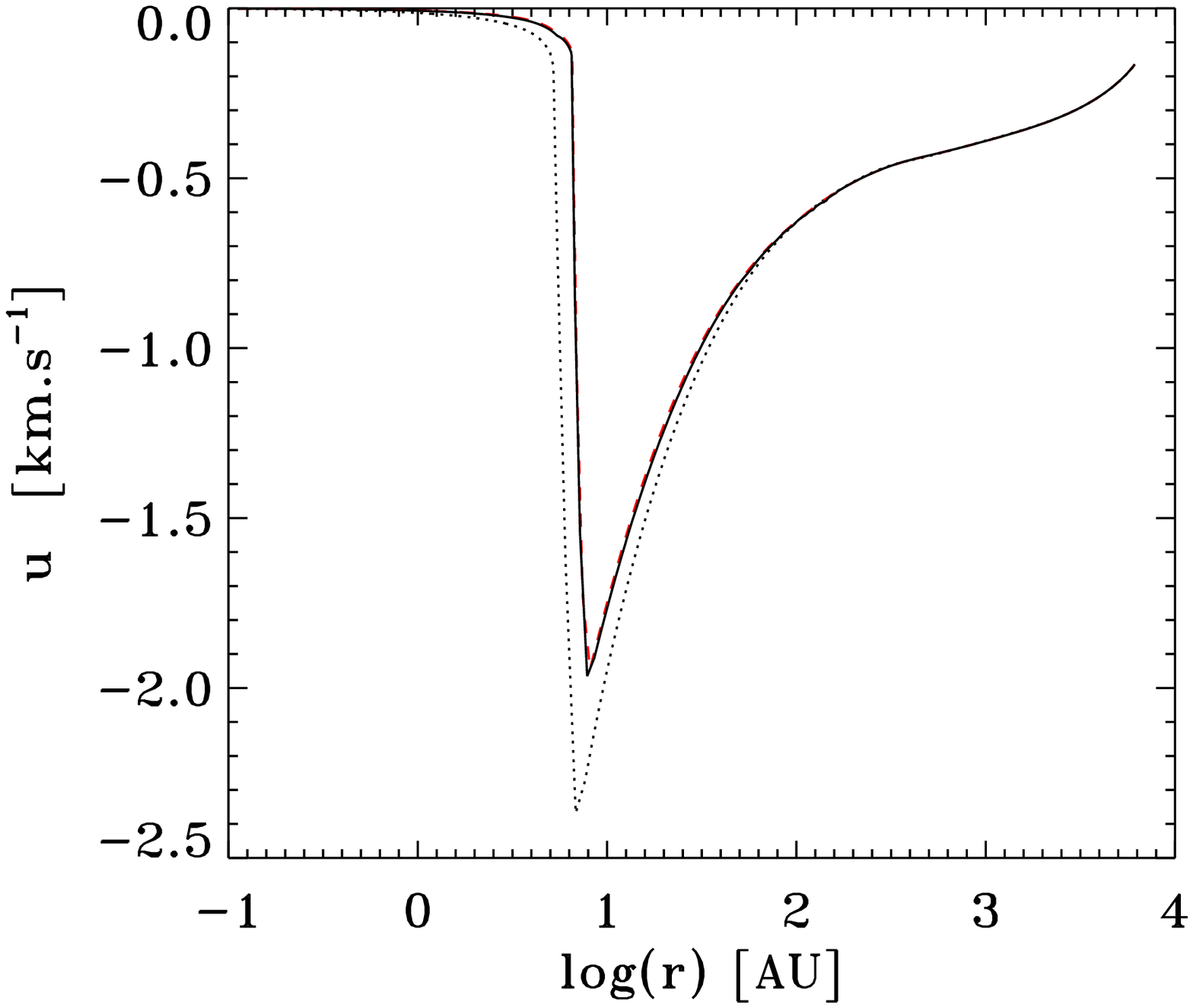}
  \includegraphics[width=6.5cm,height=5cm]{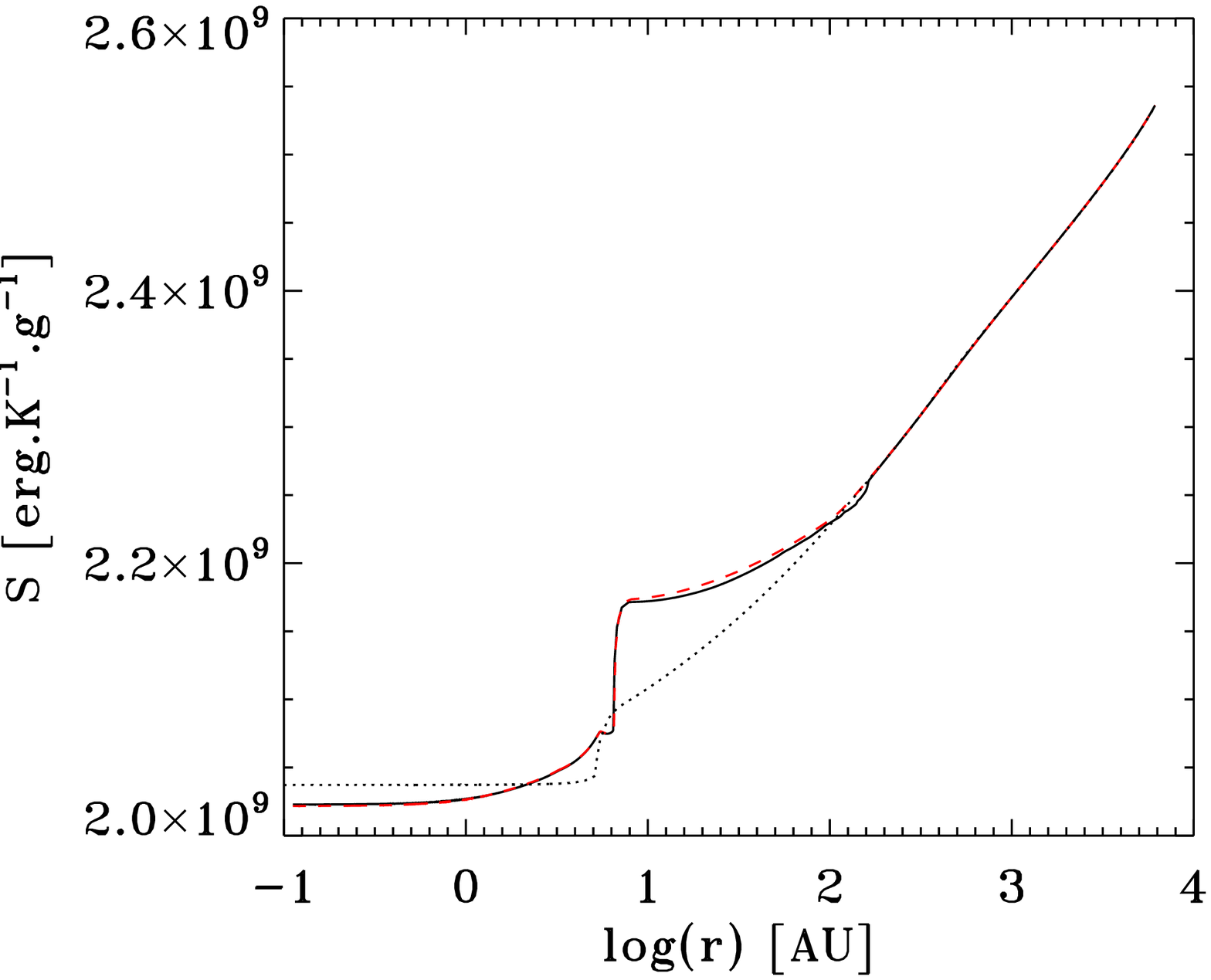}
  \includegraphics[width=6.5cm,height=5cm]{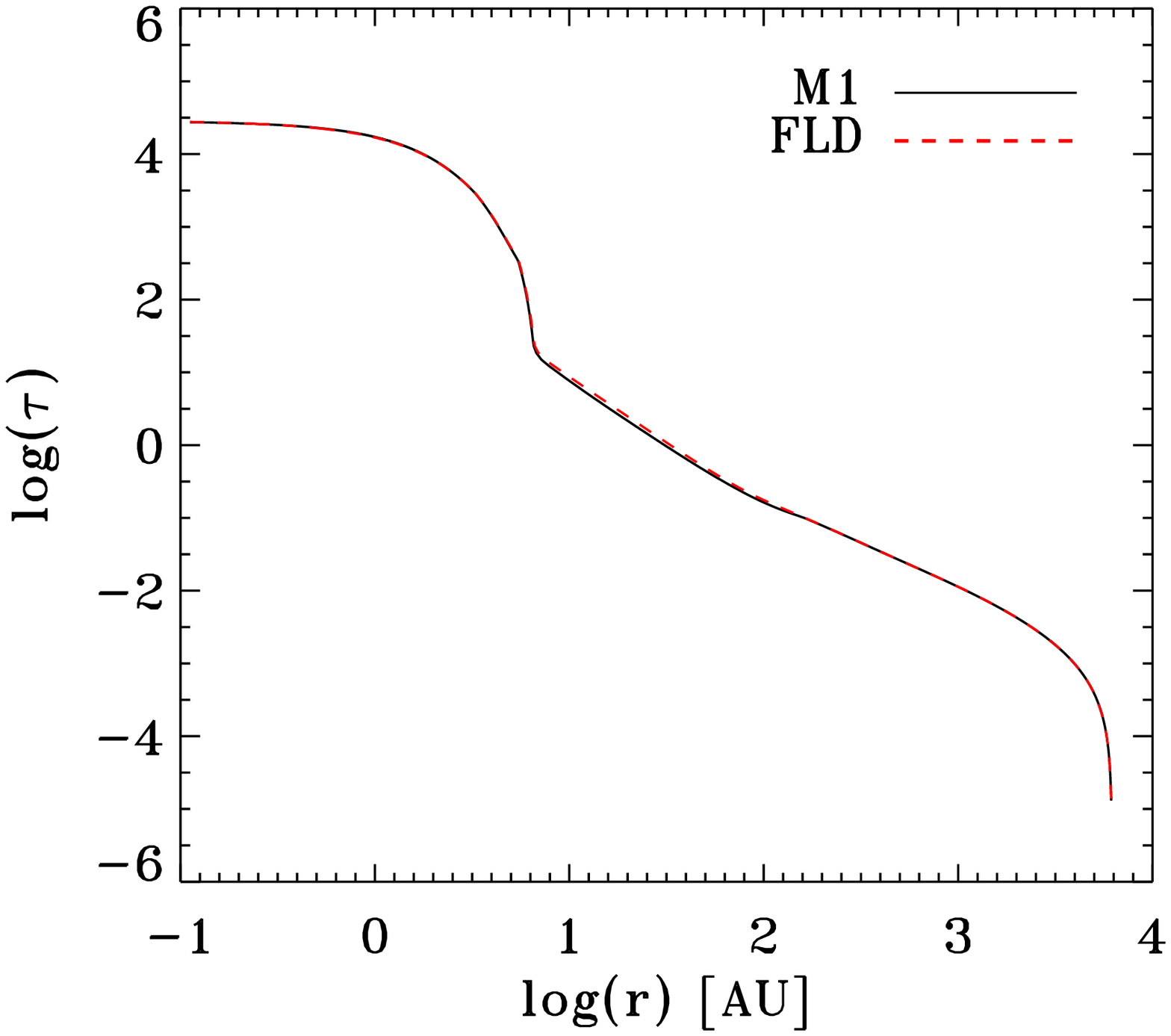}
  \includegraphics[width=6.5cm,height=5cm]{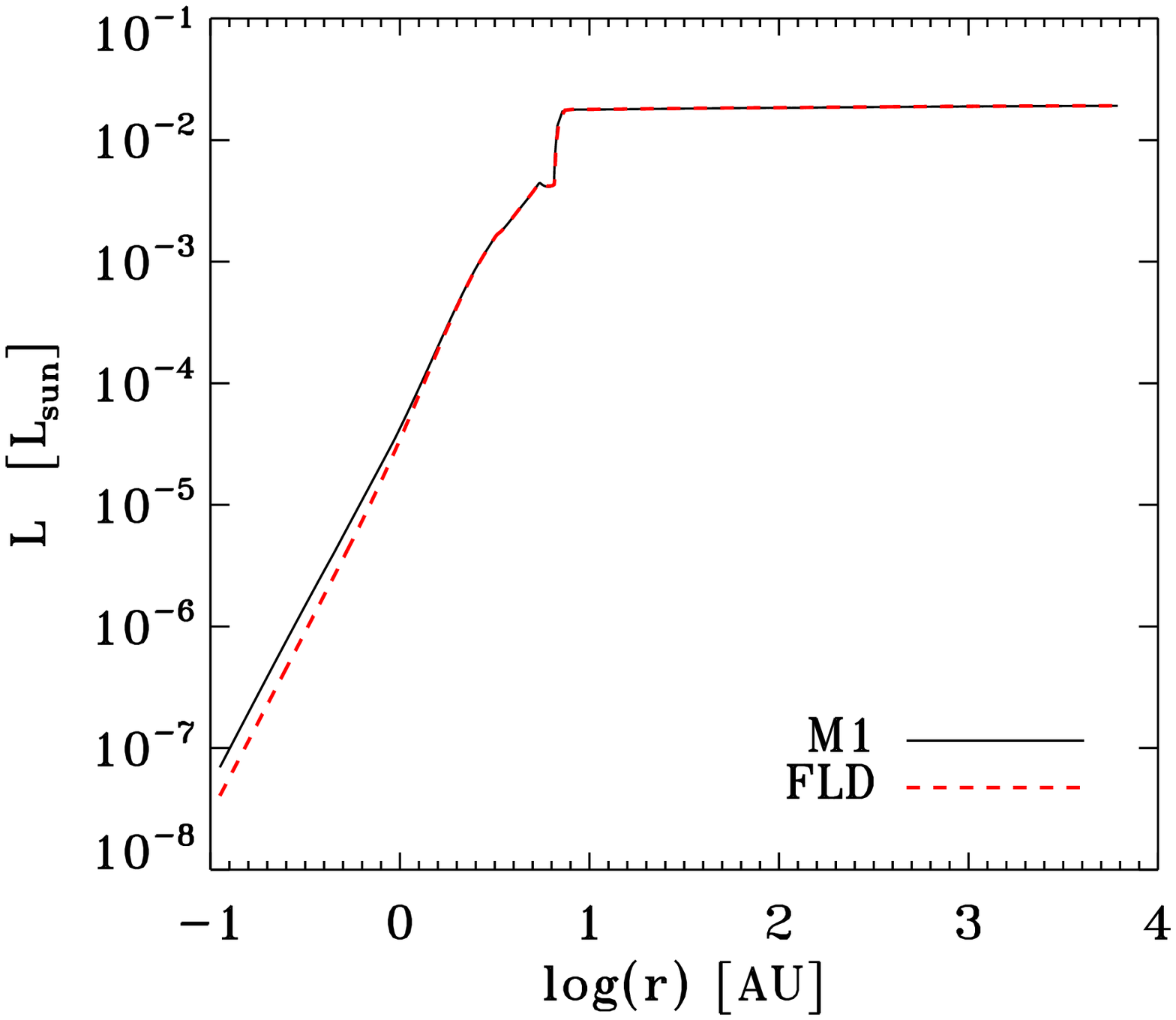}
  \includegraphics[width=6.5cm,height=5cm]{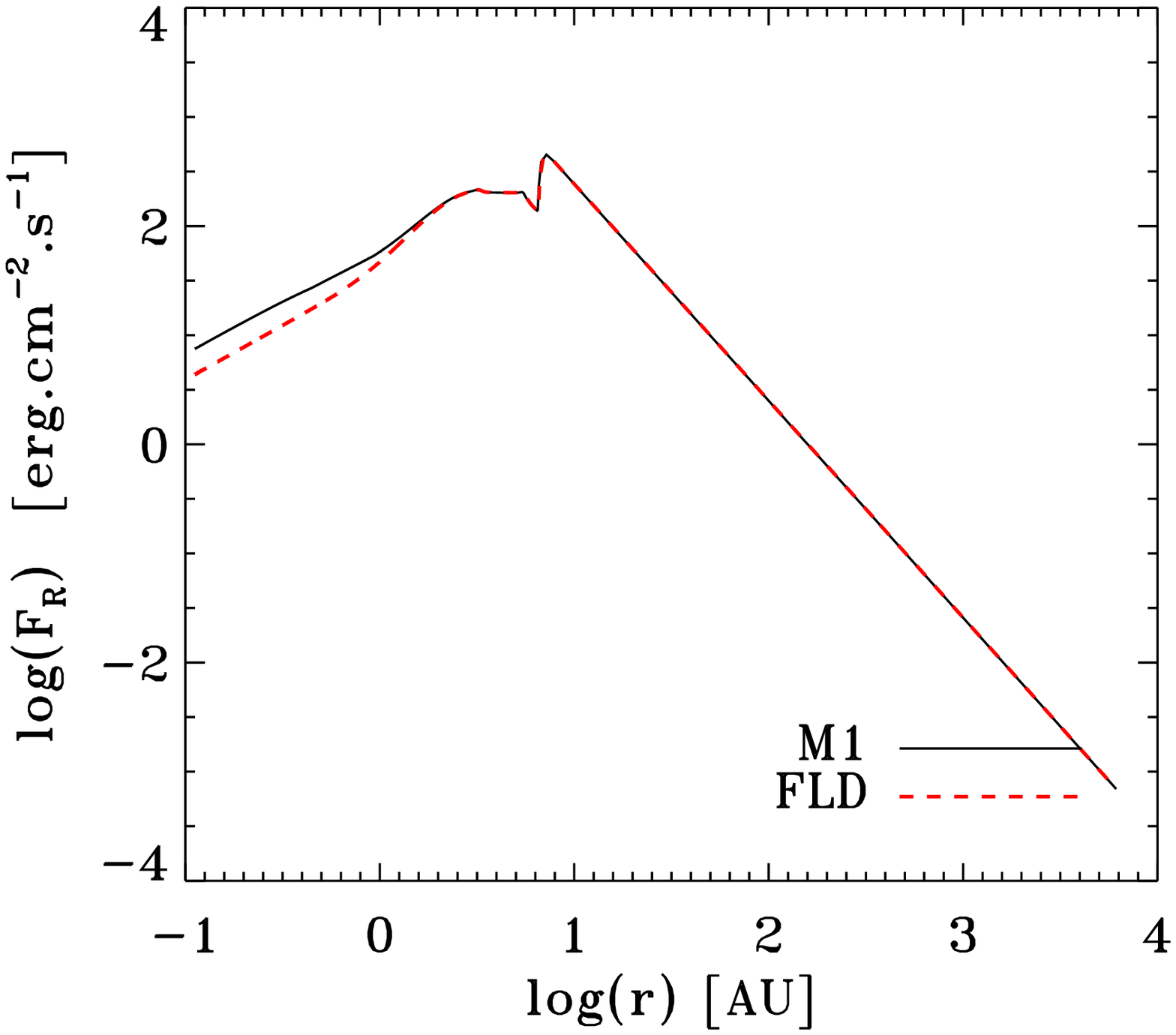}
  \includegraphics[width=6.5cm,height=5cm]{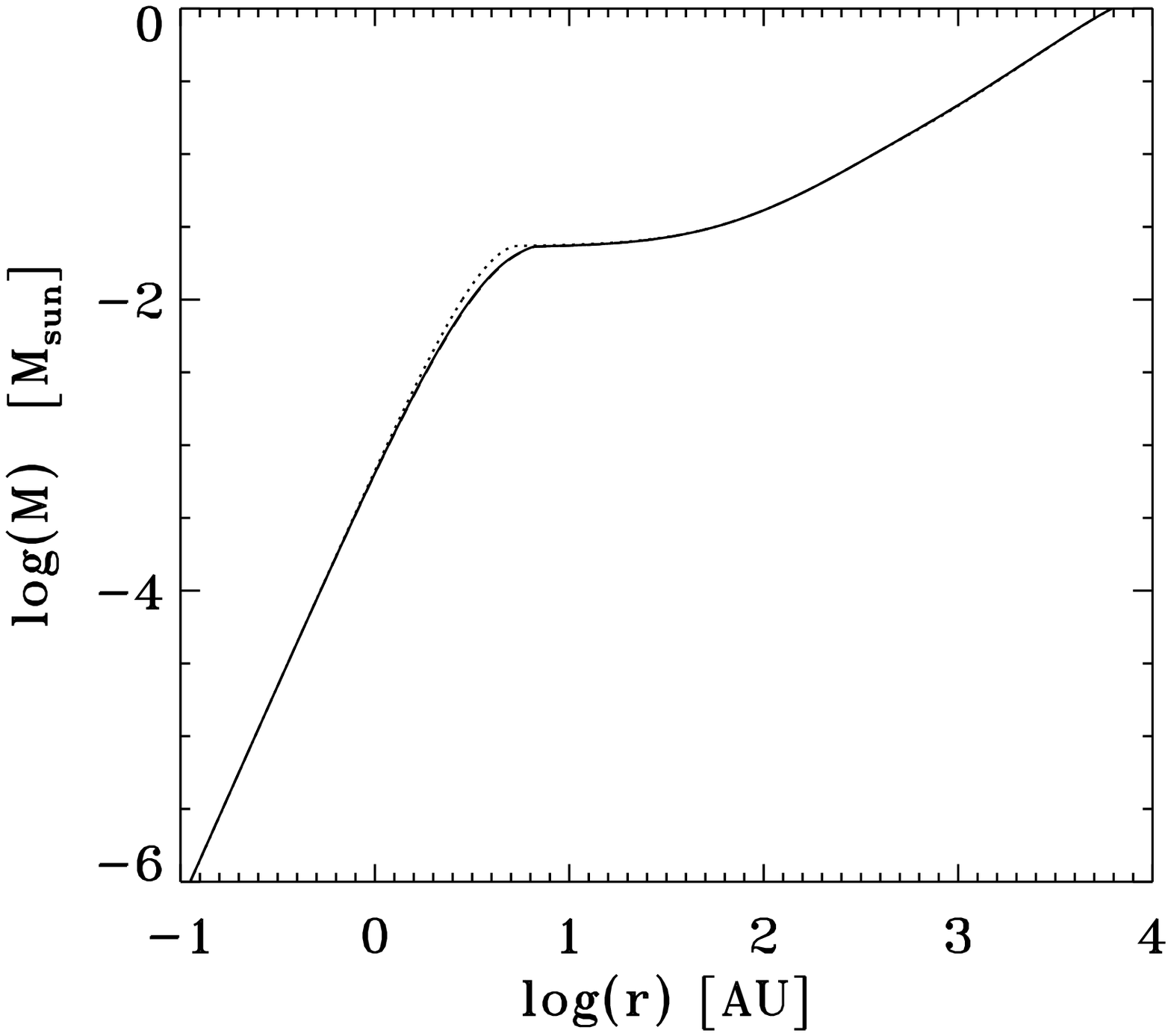}
  \caption{Radial profiles of various properties during the collapse of a 1 M$_\odot$ dense clump for a core central density $\rho_c=1\times10^{-10}$ g  cm$^{  -3}$, for  the M1 (solid line),  FLD (dashed red line) and  barotropic (dashed-dotted line) models.  From top left to bottom right: ({\it a})  density, ({\it b}) gas temperature, ({\it
    c}) velocity, ({\it d}) entropy,  ({\it e}) optical depth, ({\it f})
  luminosity, ({\it g}) radiative flux and ({\it h}) integrated mass.}
\label{plot_dc_1dm10}
\end{figure*}

\subsection{Mechanical and thermal profile of the first prestellar core.}
	
Figure  \ref{plot_dc_1dm10} displays  the radial profiles of  the density, gas
temperature, velocity, specific entropy, optical depth, luminosity, radiative
flux  and integrated  mass,  once the central
density  reaches $\rho_c=1\times10^{-10}$ g cm$^{  -3}$, for  calculations with
M1, FLD and  the barotropic EOS.  As mentioned  before, the first core
radius is smaller  with the barotropic model, whereas  both M1 and FLD
results  are  similar. For  the  latter  models,  we find  only  small
differences  around $\tau\sim  1$,  i.e.  at the  transition between  the
optically  thick  and thin  regimes.   Since the  M1 model  naturally
recovers the diffusion and  transport regimes and since the FLD  model is defined as
to recover these limits as well, it is natural that we get
similar  results   when  either the  transport  or the diffusion regimes  are  well
established.   Although  the  accretion   shock  is  located within the
transition region around $\tau \sim  10$,  the aforementioned small
differences do not  affect the first core properties. The  entropy jump at
the accretion shock  is much higher with M1 and FLD than with the barotropic approximation.  We also see from
the temperature  and entropy profiles that the  barotropic  EOS cannot handle 
correctly the transition  from the  isothermal to the adiabatic regime, as pointed out earlier.
The radiative precursor  in front of the shock  is not reproduced with
the barotropic  EOS. In that case, the gas becomes rapidly isothermal  after the
shock, whereas it  is heated by photons escaping  from the shock in the 
FLD and M1 models. As a  consequence, the core temperature is higher in the  barotropic  case.  The differences in  the  behavior of  the
radiative  flux  at  small   radius  come  from  the  differences  in
temperature between the M1 and  FLD models. 

The emergent luminosity, on the other hand, is the same  for both the M1 and FLD models.  The luminosity jump, $\sim 0.013$ L$_\odot$
is  consistent  with the  accretion  luminosity  estimated from  equation
(\ref{Lacc}),  $0.014$  L$_\odot$.   This means  that  all  the
infalling  kinetic  energy  is   radiated  away at the shock boundary,  i.e. that the  shock  is
supercritical at the formation of the first core.

If  we  apply  the  criterion  derived in  equations  (\ref{Tcr})  and
(\ref{Tcr2})      to  the   upstream     quantities      estimated     in
Fig.  \ref{plot_dc_1dm10}, i.e. $  \rho_1   \sim  7  \times  10^{-14}$  g
cm$^{-3}$, $u_1 \sim 2 \times10^5$cm s$^{-1}$, and $T_1\sim 57$ K, the
corresponding critical  temperatures are $T_\mathrm{cr}\sim  23$ K with
(\ref{Tcr})  and  $T_\mathrm{cr}\sim  57$  K with  (\ref{Tcr2}).  This
confirms the fact that the accretion shock  on the first core
is  supercritical. However,  since accretion takes
place  in  the transition  region  between  optically  thin and  thick
regimes and since  most of the upstream region  is optically thin (see
luminosity and optical  depth profiles in Fig. \ref{plot_dc_1dm10}),
the  most relevant model is the one we developed  for  a supercritical  shock with  a
upstream optically thin material.

\subsection{Comparison with an analytical model}

In Sect. 2, we have  developed   a semi-analytical solver  that can be
applied to  the accretion radiative shock  on the first  core and that
covers three  cases: the  case of sub-  or super-critical shocks  in an
optically thick material  and the case of a  supercritical shock in an
optically thin material. In the following, we develop a simple model for the protostellar
collapse, where the upstream quantities are estimated under some basic assumptions, and we compare 
this model with our numerical results. We need to estimate  the density,  the  velocity and  the
temperature in  the preshock region in  the context of  the first core
formation.   We  can  easily  get  the  density  (and  the velocity)  from
the Larson-Penston  and Shu self-similar  solutions.  The tricky  part
is to estimate the temperature before the shock. However,
as mentioned in \S\ref{sec42},  the preshock temperature is determined by
the upstream  velocity. Assuming that the accretion shock  is supercritical
and that the  upstream material is optically thin, it  is then easy to
get     this    temperature     and    to     recover     all    fluid
variables. \cite{Omukai_07}  proposes an alternative  model that fully
describes the  first core characteristics  but does not  use jump
relations for  a radiating fluid. In our model, we  consider the
characteristics at the first core border and the jump conditions for a radiating fluid.  \\

We approximate the upstream velocity by the free-fall velocity
\begin{equation}
u_\mathrm{1,ff}=\sqrt{\frac{2 G M_\mathrm{fc}}{R_\mathrm{fc}} }.
\label{u1}
\end{equation}

The  preshock density is given  assuming  a profile $\propto r^{-2}$  in the
free-falling envelope
\begin{equation}
\rho_1=\frac{c_s^2}{2\pi G}R_\mathrm{fc}^{-2},
\label{d1}
\end{equation}
where $c_s=(P/\rho)^{1/2}$ is the isothermal sound speed (at 10 K).  The
temperature  is estimated by  assuming a  supercritical shock,  i.e.  all the
upstream kinetic  energy being radiated away.  Moreover, our calculations
using the FLD or M1  models yield $\mathcal{M}_1  \sim 2$  and
Fig. \ref{PX_fM} shows that $X  \sim 1$. The shock
temperature thus reads (c.f. equation \ref{Tp_super})
\begin{equation}
\sigma T_\mathrm{s}^4=\frac{\rho_1 u_{1,\mathrm{ff}}^3}{2}.
\label{T1}
\end{equation}

We take the first core properties obtained in our numerical calculations,
similar to those in \cite{Masunaga_Miyama_Inutsuka_I_1998ApJ}, i.e. $M_\mathrm{fc} \sim 2.3 \times  10^{-2}$ M$_\odot$ and $R_\mathrm{fc}
\sim  8   $  AU. If we apply equations (\ref{u1}),   (\ref{d1})  and  (\ref{T1}) to these quantities,
we get:   $\rho_1=5.8\times10^{-14}$  g  cm$^{-3}$,
$u_1\sim2.3 \times10^5$  cm s$^{-1}$  and $T_1\sim50$  K. 

We now apply these preschock quantities to the jump properties derived in Sect. \ref{model} for a supercritical shock with an optically thin upstream material. 
According to eq. \ref{rPi}, we get $r=\rho_2/\rho_1\sim50$, which gives $\rho_2\sim 2.9\times 10^{-12}$ g cm$^{-3}$, and $u_2=4.6\times10^{3}$ cm s$^{-1}$.
Using  equation (\ref{XX}),   we   get   $X\sim   0.99$: all the infalling kinetic energy is radiated away at the shock.

We can compare these analytical results with the numerical ones, given  in   Fig.
\ref{plot_dc_1dm10}  and  in   Table  \ref{M1_DIFF}:
$\rho_1\sim7\times10^{-14}$  g cm$^{-3}$, $u_1\sim2\times10^5$  cm s$^{-1}$
and $T_1=57$ K. Thus, the analytical  estimates for the preshock quantities agree quite well with the  numerical values.  We also read from the figure and table
$\rho_2\sim 2.15 \times10^{-12}$   g    cm$^{-3}$,   $u_2 \sim 10^4$   cm
s$^{-1}\ll u_1$, again very similar  to our analytical estimates. These comparisons validate our simple analytical model to infer the properties of the first Larson's core. \\

\section{Summary and perspectives}

In this paper,  we have conducted RHD calculations aimed at describing the physical and radiative properties of radiative shocks. We have applied these calculations to the formation of the first Larson's core, during the early phases of protostellar collapse. We have also derived simple analytical models and compared the results with the ones obtained in the simulations. The main results of our study can be summarized as follows :

\begin{enumerate}
\item The properties  of the first collapse and the characteristics of the
first   core in our calculations are   in   good    agreement   with the ones obtained by
\cite{Masunaga_Miyama_Inutsuka_I_1998ApJ}.   We  find  that the  first
core has  a typical radius of $\sim  8$ AU and a  mass $\sim2 \times 10^{-2}$
M$_\odot$. This sets up the initial conditions for the second collapse and the formation of the second Larson's core. 

\item We show that, at the first core stage,
the accretion shock is a supercritical radiative shock, at which all
the infalling kinetic energy is radiated away, and that a barotropic EOS cannot reproduce the
correct jump conditions at the shock. The FLD and M1 calculations show that
there is a substantial entropy loss during the formation of the first core, due to the
radiative loss. A barotropic EOS cannot handle  correctly this cooling mechanism.  In consequence, the
first core's entropy content obtained with a barotropic approximation is overestimated, compared with
the calculations which solve the radiative transfer.
Such a cooling effect can have a strong impact on the core fragmentation process and, eventually on the initial properties of the future protostar (Commer\c con et al., {\it in prep}).

\item We confirm that, when radiative cooling is properly taken into account, the transition from an isothermal to an adiabatic phase during the first collapse and the formation of the first Larson's core does not necessarily correspond to an optical depth of unity, as shown initially by Masunaga \& Inutsuka (1999).

\item 
We develop a simple
analytical model for supercritical shocks within an optically thin medium,
which reproduces well the jump  conditions obtained with the numerical calculations. We
show that the compression ratio in such a kind of shock can become very
high  ($r\sim50$).  We plan in the future to keep exploring this issue with  a  frequency  dependent radiative transfer model.
Indeed, strong shocks  on (massive)
protostars are  known to  be optically thick  for hard  photons, while
optically thin for UV radiation \citep[e.g. ][]{SST}.

\item We  show that, at least  in 1D spherical  calculations, the flux
limited diffusion approximation is appropriate to  study the earliest stages of star
formation, as it gives  very similar results as the more complete  M1 model for radiative transfer.  Note, however,
that our 1D  spherical geometry code  cannot account for  multi dimensional
effects like the anisotropy of the radiation field.  
\end{enumerate}

This study confirms the necessity to solve the complete RHD equations, i.e. to correctly take into account the coupling between radiation and hydrodynamics, when addressing strong shock conditions, as occurring during the formation of the first Larson's core in the context of star formation. Such a complete, consistent treatment of radiation and hydrodynamics is necessary to correctly handle the cooling properties of the accreting gas and thus to obtain the correct mechanical and thermal properties of the first core. This, in turn, sets up the initial conditions for the second collapse and the formation of the second core. Indeed, since  the first core has a  short lifetime (a
few  hundred  years),  it is important to pursue these calculations during the second  collapse, including H$_2$ dissociation. Such work is under progress.
\begin{acknowledgements}
Calculations have been performed on the GODUNOV cluster at SAp/CEA. The research of BC is granted by  the postdoctoral fellowships from the Max-Planck-Institut f\"{u}r Astronomie.
The research leading to these results has received funding from the European Research Council under the European Community's Seventh Framework Programme (FP7/2007-2013 Grant Agreement no. 247060)
\end{acknowledgements}
\bibliographystyle{aa}
\bibliography{biblio}

\begin{thebibliography}{30}
\expandafter\ifx\csname natexlab\endcsname\relax\def\natexlab#1{#1}\fi

\bibitem[{{Audit} {et~al.}(2002){Audit}, {Charrier}, {Chi{\`e}ze}, \&
  {Dubroca}}]{Audit_et_al_2002}
{Audit}, E., {Charrier}, P., {Chi{\`e}ze}, J.~P., \& {Dubroca}, B. 2002, ArXiv
  Astrophysics e-prints

\bibitem[{{Bouquet} {et~al.}(2000){Bouquet}, {Teyssier}, \&
  {Chi{\`e}ze}}]{Bouquet_2000}
{Bouquet}, S., {Teyssier}, R., \& {Chi{\`e}ze}, J.~P. 2000, \apjs, 127, 245

\bibitem[{{Calvet} \& {Gullbring}(1998)}]{Calvet_1998}
{Calvet}, N. \& {Gullbring}, E. 1998, \apj, 509, 802

\bibitem[{{Commer{\c c}on} {et~al.}(2008){Commer{\c c}on}, {Hennebelle},
  {Audit}, {Chabrier}, \& {Teyssier}}]{Commercon_2008}
{Commer{\c c}on}, B., {Hennebelle}, P., {Audit}, E., {Chabrier}, G., \&
  {Teyssier}, R. 2008, \aap, 482, 371

\bibitem[{{Commer{\c c}on} {et~al.}(2010){Commer{\c c}on}, {Hennebelle},
  {Audit}, {Chabrier}, \& {Teyssier}}]{Commercon_2010L}
{Commer{\c c}on}, B., {Hennebelle}, P., {Audit}, E., {Chabrier}, G., \&
  {Teyssier}, R. 2010, \aap, 510, L3+

\bibitem[{{Commer{\c c}on} {et~al.}(2011){Commer{\c c}on}, {Teyssier}, {Audit},
  {Hennebelle}, \& {Chabrier}}]{Commercon_2011}
{Commer{\c c}on}, B., {Teyssier}, R., {Audit}, E., {Hennebelle}, P., \&
  {Chabrier}, G. 2011, ArXiv e-prints

\bibitem[{{Drake}(2005)}]{Drake_2005}
{Drake}, R.~P. 2005, \apss, 298, 49

\bibitem[{Drake(2007)}]{Drake_2007}
Drake, R.~P. 2007, Physics of Plasmas, 14, 043301

\bibitem[{{Dubroca} \& {Feugeas}(1999)}]{Dubroca_1999}
{Dubroca}, B. \& {Feugeas}, J.~N. 1999, Comptes Rendus de l'Acadmie des
  Sciences, 329, 915

\bibitem[{{Ferguson} {et~al.}(2005){Ferguson}, {Alexander}, {Allard}, {Barman},
  {Bodnarik}, {Hauschildt}, {Heffner-Wong}, \& {Tamanai}}]{Ferguson_05}
{Ferguson}, J.~W., {Alexander}, D.~R., {Allard}, F., {et~al.} 2005, \apj, 623,
  585

\bibitem[{{Gonz{\'a}lez} {et~al.}(2007){Gonz{\'a}lez}, {Audit}, \&
  {Huynh}}]{Gonzalez_2007}
{Gonz{\'a}lez}, M., {Audit}, E., \& {Huynh}, P. 2007, \aap, 464, 429

\bibitem[{{Gonz{\'a}lez} {et~al.}(2009){Gonz{\'a}lez}, {Audit}, \&
  {Stehl{\'e}}}]{Gonzalez_2009}
{Gonz{\'a}lez}, M., {Audit}, E., \& {Stehl{\'e}}, C. 2009, \aap, 497, 27

\bibitem[{{Larson}(1969)}]{Larson_1969}
{Larson}, R.~B. 1969, \mnras, 145, 271

\bibitem[{{Levermore}(1984)}]{Levermore_1984JQSRT}
{Levermore}, C.~D. 1984, Journal of Quantitative Spectroscopy and Radiative
  Transfer, 31, 149

\bibitem[{{Masunaga} \& {Inutsuka}(2000)}]{Masunaga_Inutsuka_2000}
{Masunaga}, H. \& {Inutsuka}, S.-i. 2000, \apj, 531, 350

\bibitem[{{Masunaga} {et~al.}(1998){Masunaga}, {Miyama}, \&
  {Inutsuka}}]{Masunaga_Miyama_Inutsuka_I_1998ApJ}
{Masunaga}, H., {Miyama}, S.~M., \& {Inutsuka}, S.-I. 1998, \apj, 495, 346

\bibitem[{{Michaut} {et~al.}(2009){Michaut}, {Falize}, {Cavet}, {Bouquet},
  {Koenig}, {Vinci}, {Reighard}, \& {Drake}}]{Michaut_2009}
{Michaut}, C., {Falize}, E., {Cavet}, C., {et~al.} 2009, \apss, 322, 77

\bibitem[{{Mihalas} \& {Mihalas}(1984)}]{Mihalas_book}
{Mihalas}, D. \& {Mihalas}, B.~W. 1984, {Foundations of radiation
  hydrodynamics}, ed. D.~{Mihalas} \& B.~W. {Mihalas}

\bibitem[{{Minerbo}(1978)}]{Minerbo_1978JQSRT}
{Minerbo}, G.~N. 1978, Journal of Quantitative Spectroscopy and Radiative
  Transfer, 20, 541

\bibitem[{{Omukai}(2007)}]{Omukai_07}
{Omukai}, K. 2007, \pasj, 59, 589

\bibitem[{{Penston}(1969)}]{Penston_1969}
{Penston}, M.~V. 1969, \mnras, 144, 425

\bibitem[{{Semenov} {et~al.}(2003){Semenov}, {Henning}, {Helling}, {Ilgner}, \&
  {Sedlmayr}}]{Semenov_et_al_2003A&A}
{Semenov}, D., {Henning}, T., {Helling}, C., {Ilgner}, M., \& {Sedlmayr}, E.
  2003, \aap, 410, 611

\bibitem[{{Shu}(1977)}]{Shu_1977}
{Shu}, F.~H. 1977, \apj, 214, 488

\bibitem[{{Stahler} {et~al.}(1980){Stahler}, {Shu}, \& {Taam}}]{SST}
{Stahler}, S.~W., {Shu}, F.~H., \& {Taam}, R.~E. 1980, \apj, 241, 637

\bibitem[{{Stone} {et~al.}(1992){Stone}, {Mihalas}, \& {Norman}}]{Stone_MM_92}
{Stone}, J.~M., {Mihalas}, D., \& {Norman}, M.~L. 1992, \apjs, 80, 819

\bibitem[{{Tscharnuter}(1987)}]{tscharnuter1987}
{Tscharnuter}, W.~M. 1987, \aap, 188, 55

\bibitem[{{Tscharnuter} \& {Winkler}(1979)}]{Tscharnuter_1979}
{Tscharnuter}, W.~M. \& {Winkler}, K. 1979, Computer Physics Communications,
  18, 171

\bibitem[{{Whitworth} \& {Clarke}(1997)}]{Whitworth_Clarke_1997}
{Whitworth}, A.~P. \& {Clarke}, C.~J. 1997, \mnras, 291, 578

\bibitem[{{Winkler} \& {Newman}(1980)}]{winkler1980}
{Winkler}, K. \& {Newman}, M.~J. 1980, \apj, 236, 201

\bibitem[{{Zel'Dovich} \& {Raizer}(1967)}]{Zeldovich_book}
{Zel'Dovich}, Y.~B. \& {Raizer}, Y.~P. 1967, {Physics of shock waves and
  high-temperature hydrodynamic phenomena}, ed. Y.~B. {Zel'Dovich} \& Y.~P.
  {Raizer}

\end{thebibliography}

\begin{appendix}

\section{Effect of the numerical resolution on the energy budget through the shock. Case of a $0.01$ M$_\odot$ dense core.\label{appendixA}}

In contrast to the hydrodynamical case, the structure of a radiative shock can extend over large distances, depending on the optical properties of the material.
For an optically thin material,  the photon mean free path is large, so the shock structure is very extended compared to the viscous scale.
In this work (see Sect. \ref{num_cal}), we present numerical calculations of dense core collapse, using a fix resolution in mass, i.e. the mesh is not refined in the large gradient zones.
Although this Lagrangean description is well suited for the hydrodynamical shocks, thanks to the artificial viscosity scheme, it may encounter difficulties in the case of radiation-hydrodynamical flows, in particular in the optically thin region (upstream region, outside the first core). 

In this appendix, we present the results of the collapse of a 0.01 M$_\odot$ dense core, using the same initial ratio of thermal energy over gravitational energy as in Sec \ref{num_cal} ($\alpha\sim 0.97$). To investigate the effect of the numerical resolution, we performed calculations with 4500 cells and 18000 cells, using the FLD model. 

Figure \ref{appA1} shows the density, gas temperature, velocity, entropy, optical depth and luminosity radial profiles for the 2 calculations at a central density $\rho_c\sim1.6\times 10^{-11}$ g cm$^{-3}$. Although there are some significative differences in the radiative precursor region (i.e. the transition region between optically thin and thick regions, where $2<\tau<0.5$) and in the estimate of the first core radius ($\sim 10 \%$), the entropy, density, velocity and luminosity jumps are about the same.
In both calculations, the shock is {\it supercritical} and the amount of energy radiated away is about the same ($L=1.5\times10^{-2}$ with 4500 cells, and $L=1.44\times10^{-2}$ with 18000 cells). This means that the overall properties of the first accretion shock, including its global energy budget remain correctly calculated even at low resolution.  However, using 18000 cells, we see that the spike in the gas temperature is resolved and that the radiative precursor length is much smaller.
On the other hand, the central entropy within the first core is the same in both cases, indicating that the cooling of the first core by radiation is not affected by the lack of resolution in the radiative shock.

Figure \ref{appA2} shows the normalized energy balance at $\rho_c=1.8\times10^{-11}$ g  cm$^{  -3}$ for the calculations run with 18000 cells (left) and 4500 cells (right).
The figures display the rate of change of potential energy $dE_\mathrm{p}/dt$, kinetic energy $dE_\mathrm{k}/dt$, internal energy $dU_\mathrm{i}/dt$, total energy 
$d(E_\mathrm{k}+E_\mathrm{p}+U_\mathrm{i})/dt$, and the work done by thermal pressure  and radiative flux ($L_\mathrm{rad}+4\pi r^2Pu$). The total energy equation reads: 
\begin{equation}
\frac{d}{dt}\left( U_\mathrm{i}+E_\mathrm{k}+E_\mathrm{p}+E_\mathrm{r}\right) +\frac{\partial}{\partial r}\left[ 4\pi r^2[u(p+P_\mathrm{r})+F_\mathrm{r}]\right]=0.
\end{equation}
First, we see form fig. \ref{appA2} that the radiative pressure exerts a negligible work compared to the thermal pressure. Comparing the energy balance of the two calculations,
we see that it is globally the same, which confirms that the calculations done with 4500 cells get the correct features of the first core accretion shock, even though the radiative structure of this shock is not resolved.

\begin{figure*}
  \centering
  \includegraphics[width=6.5cm,height=5cm]{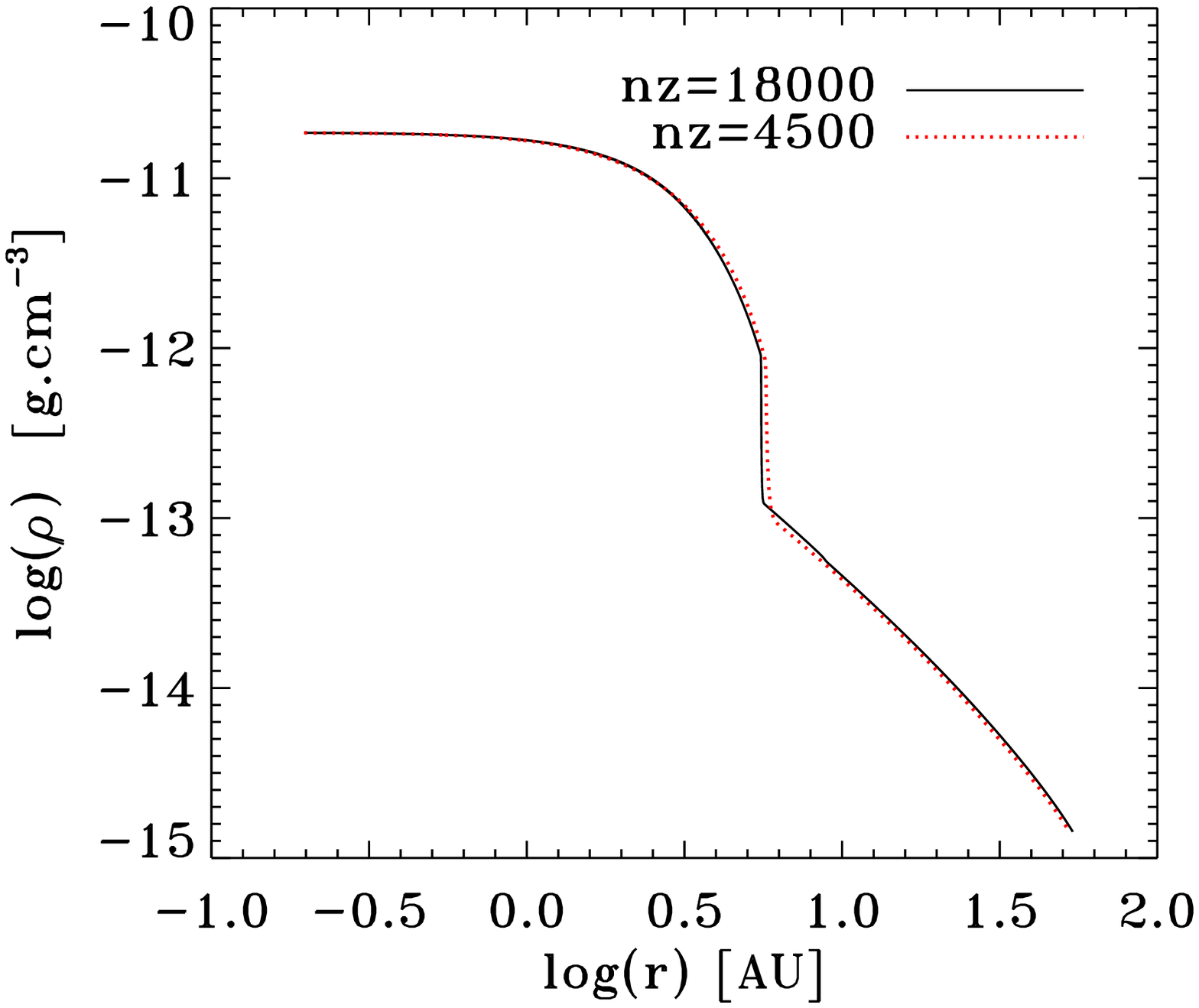}
  \includegraphics[width=6.5cm,height=5cm]{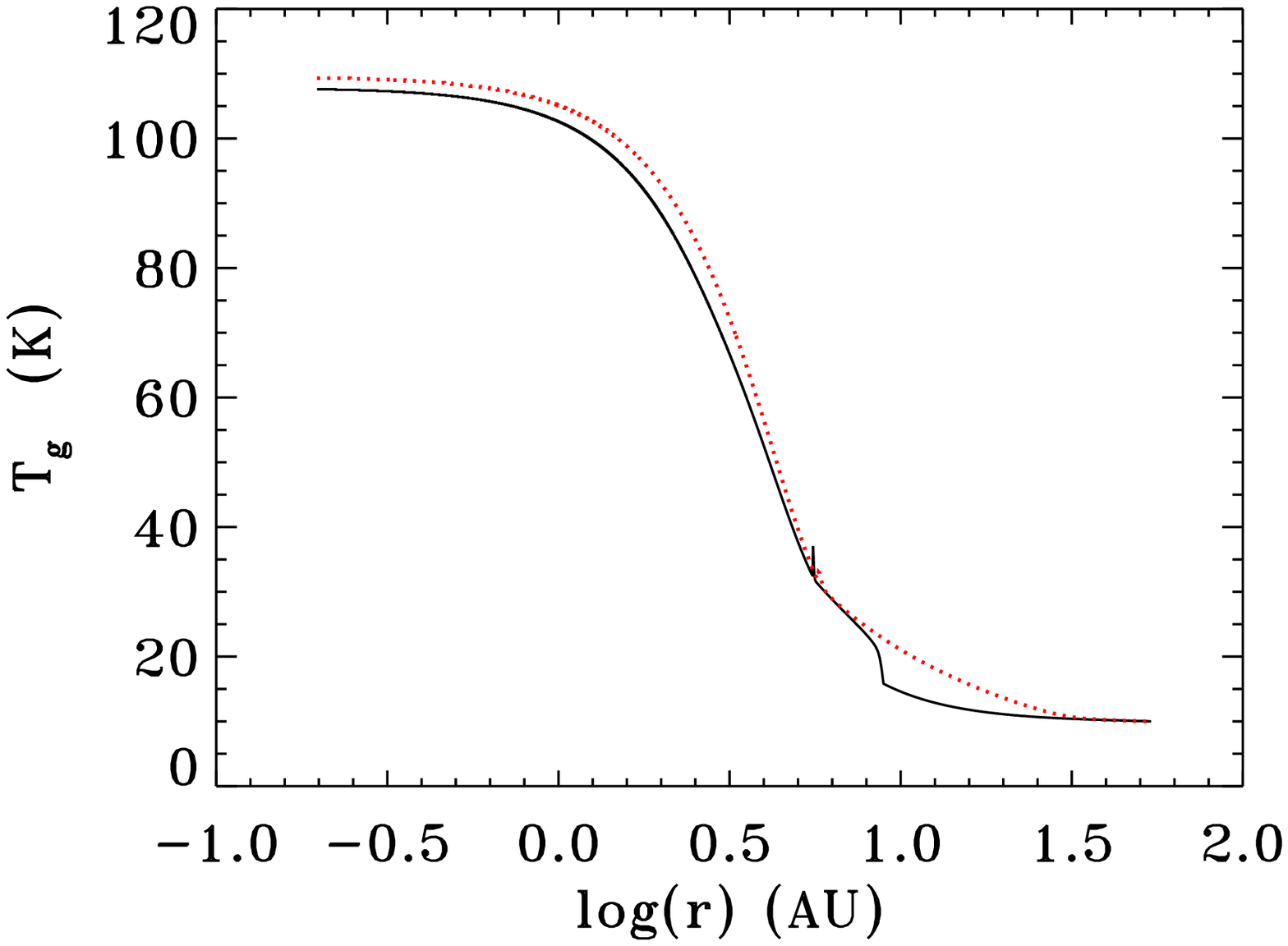}
  \includegraphics[width=6.5cm,height=5cm]{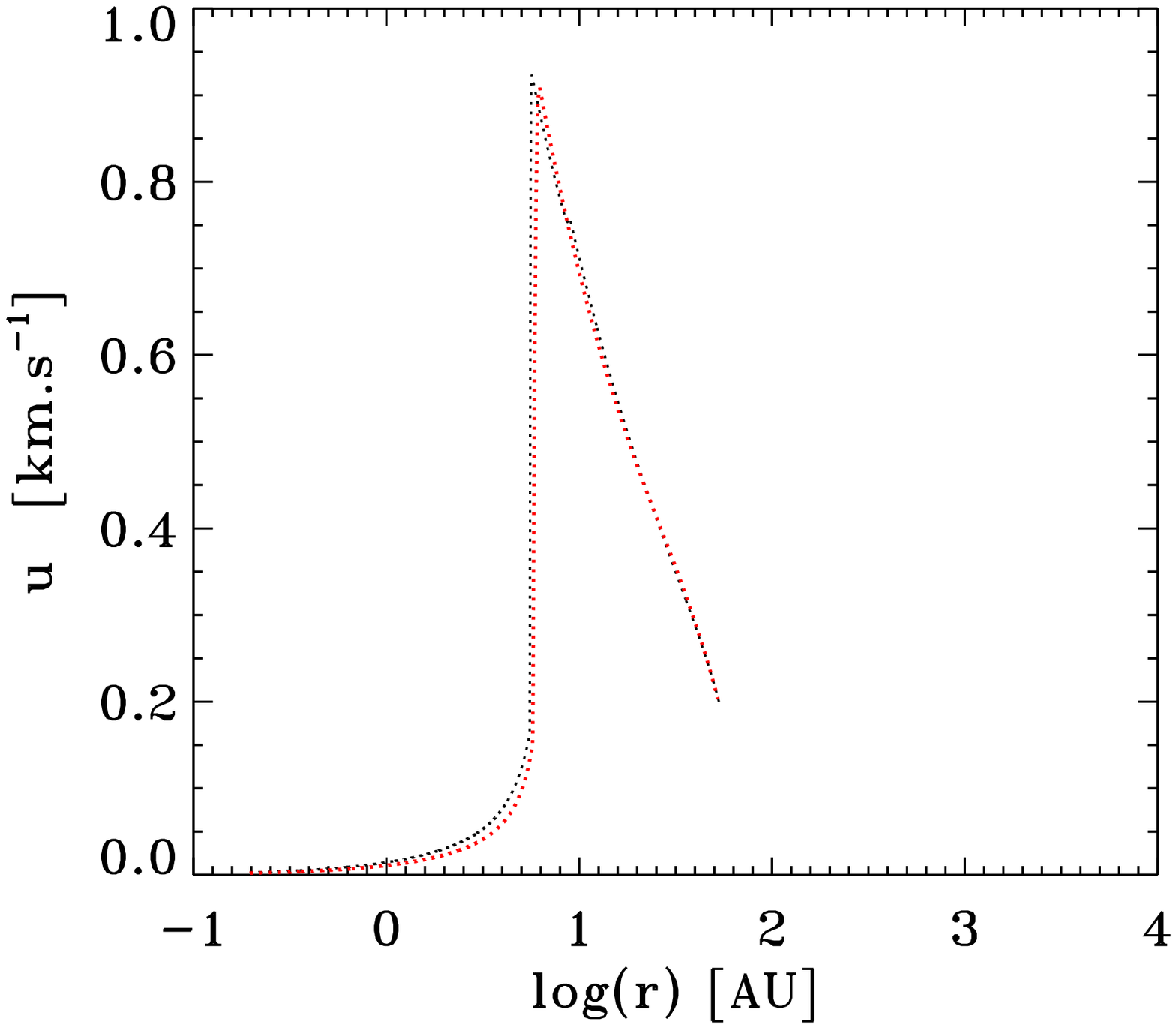}
  \includegraphics[width=6.5cm,height=5cm]{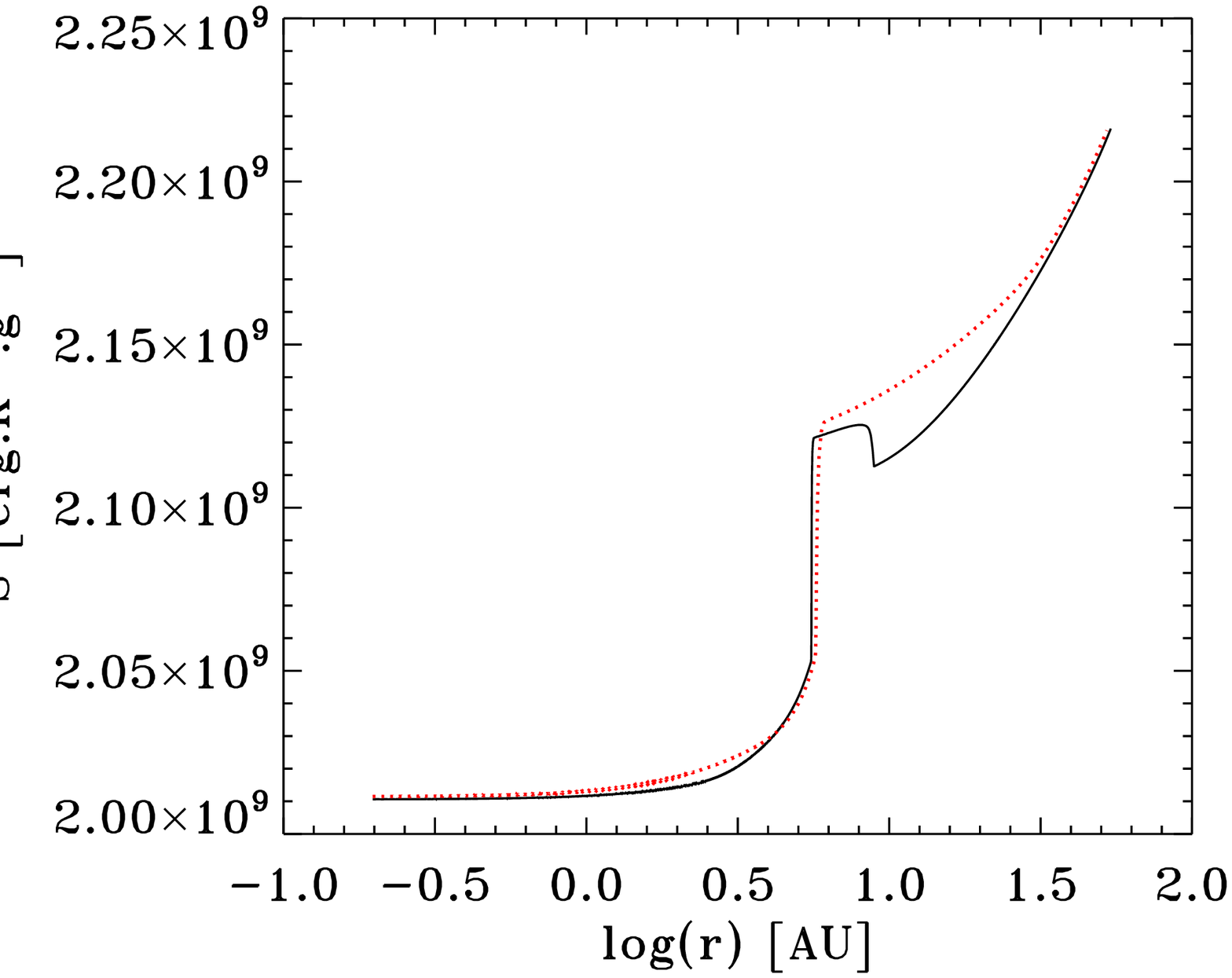}
  \includegraphics[width=6.5cm,height=5cm]{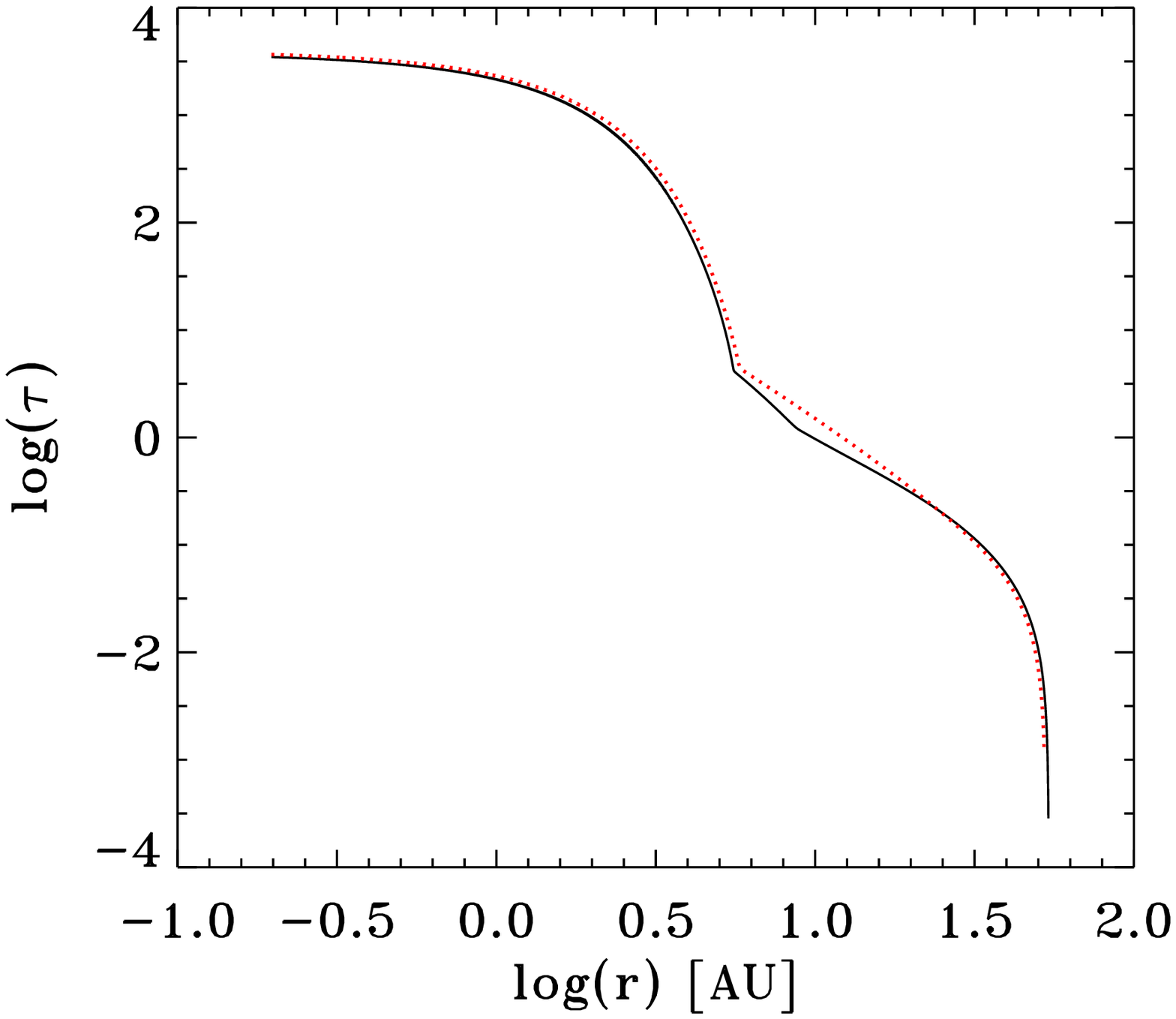}
  \includegraphics[width=6.5cm,height=5cm]{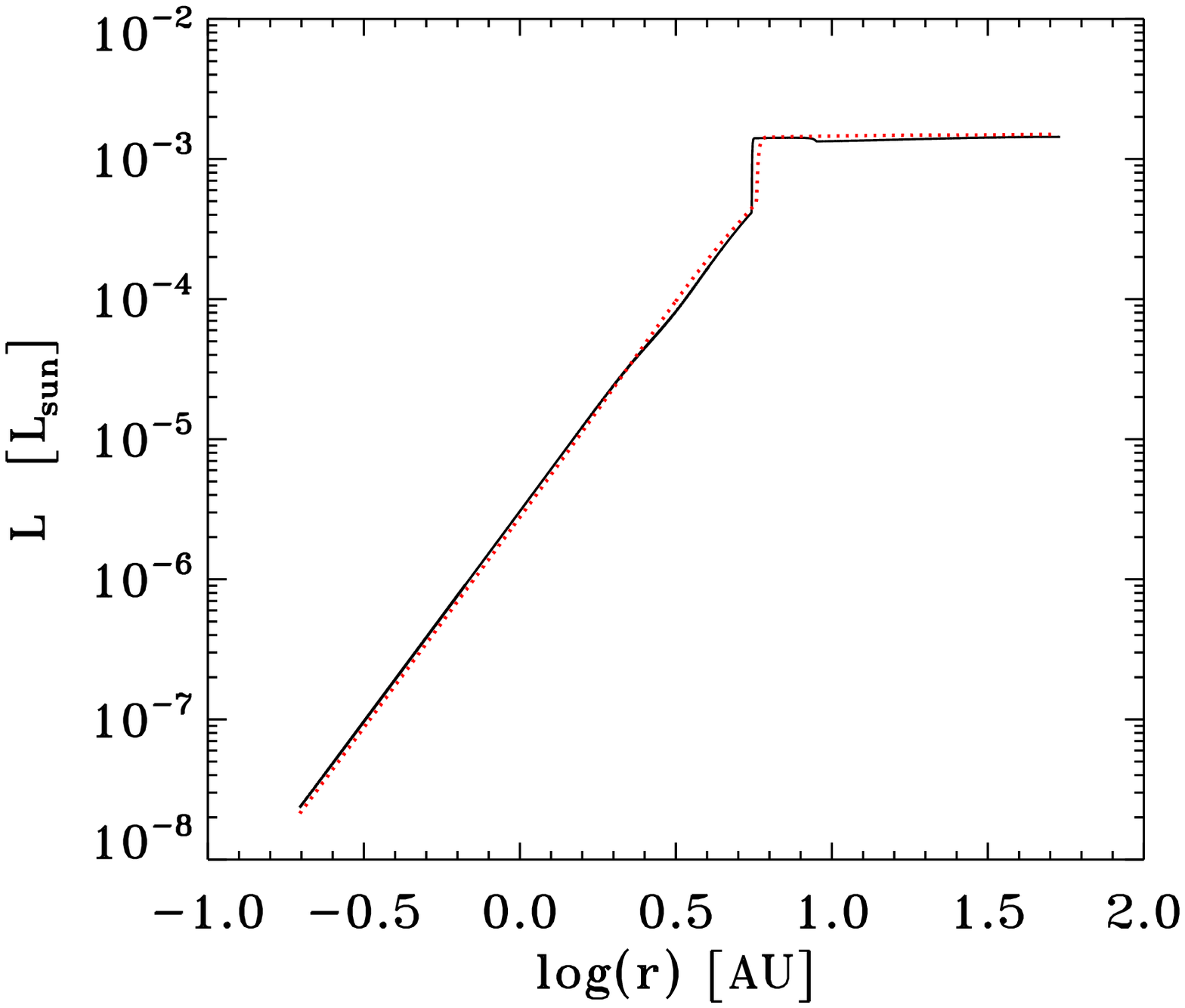}
  \caption{Radial profiles of various 1st core properties during the collapse of a 0.01 M$_\odot$ clump for a core central density $\rho_c=1.8\times10^{-11}$ g  cm$^{  -3}$, for calculations done with 4500 cells (dotted red line) and 18 000 cells (solid black line).  From top left to bottom right: ({\it a})  density, ({\it b}) gas temperature, ({\it
  c}) entropy, ({\it d}) velocity,  ({\it e}) optical depth, ({\it f})
  luminosity.}
\label{appA1}
\end{figure*}

\begin{figure*}
  \centering
  \includegraphics[width=7.5cm,height=6cm]{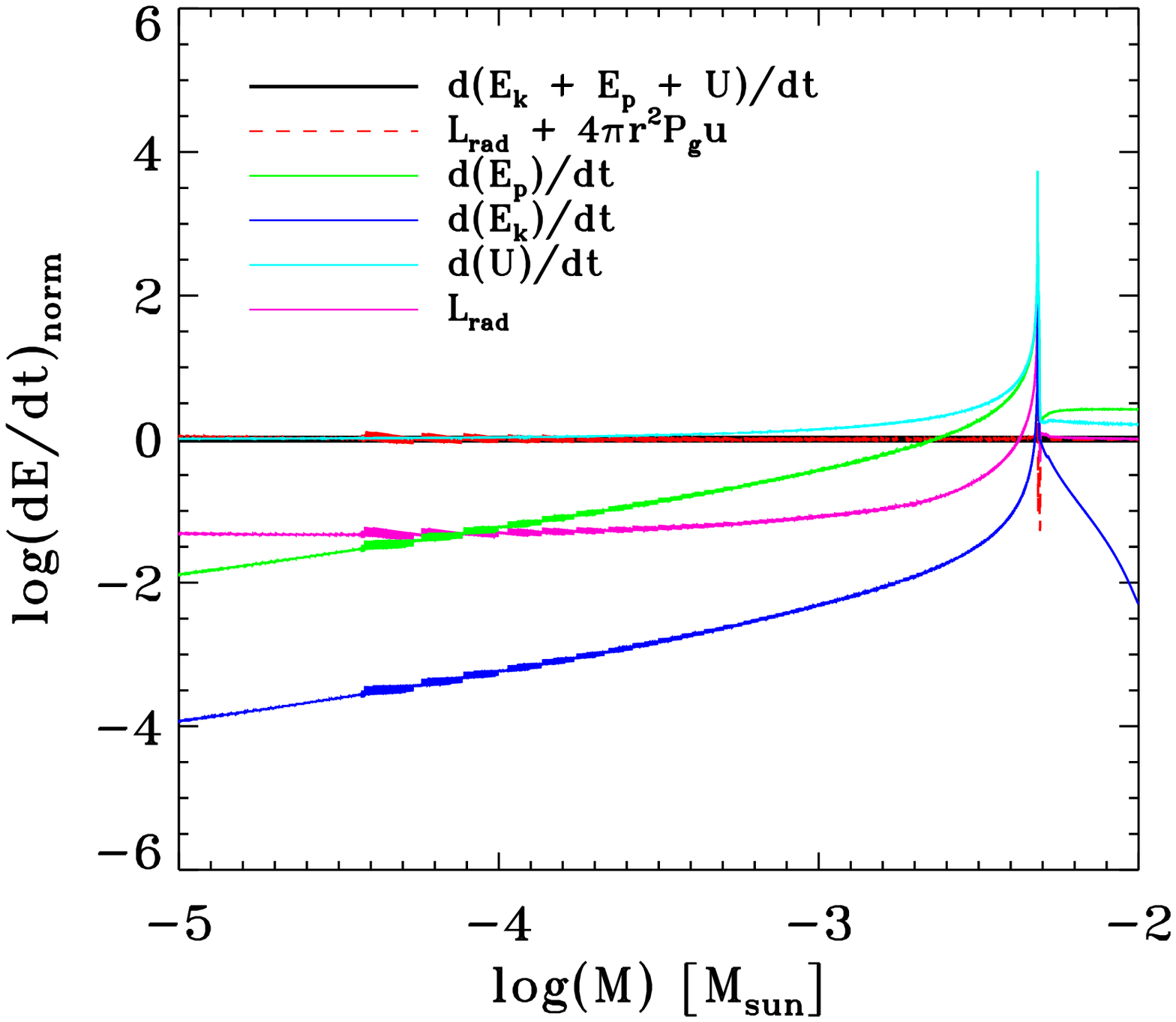}
  \includegraphics[width=7.5cm,height=6cm]{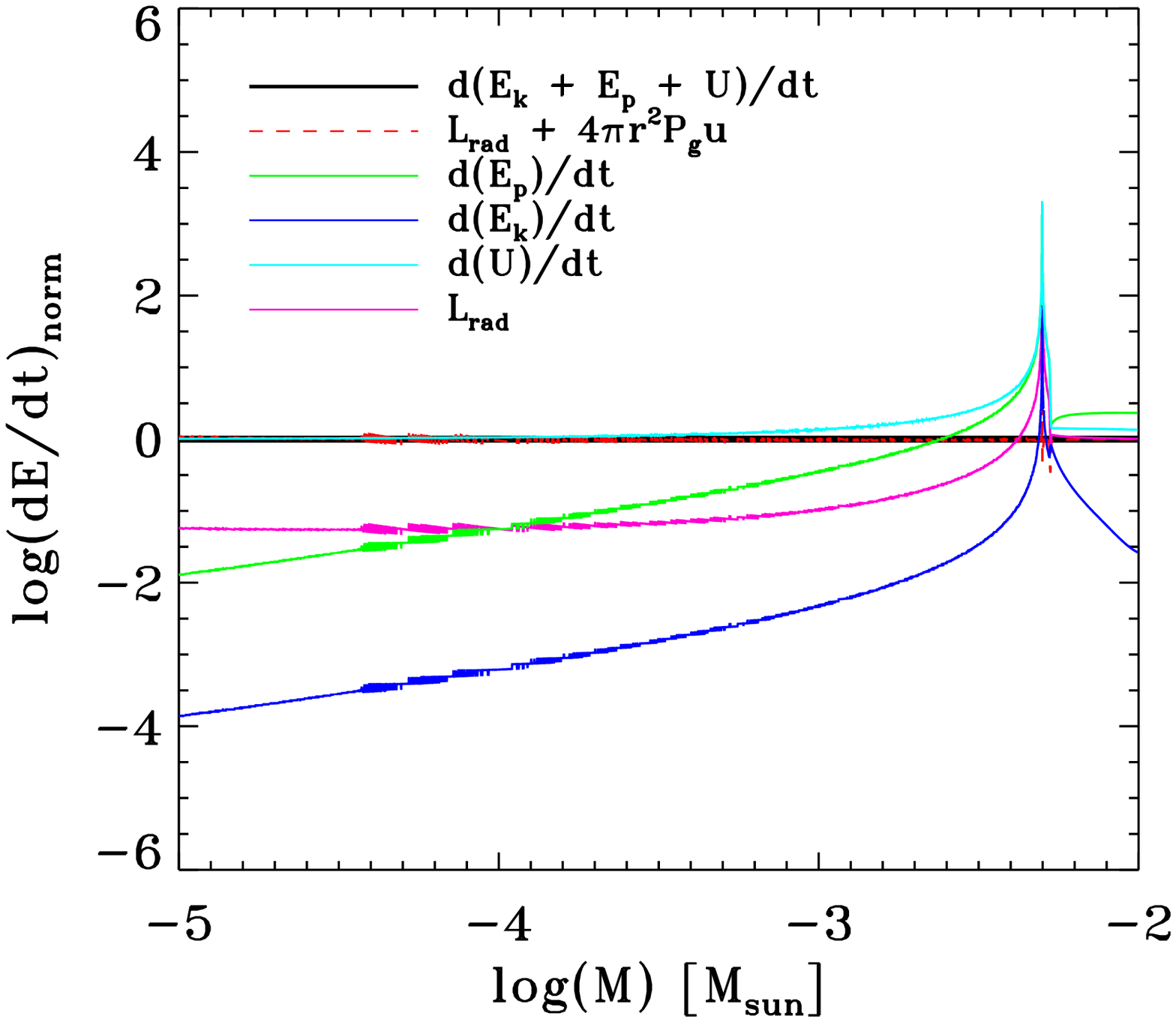}
\caption{Normalized energy balance as a function of the mass for the calculations done with 18000 cells (left) and 4500 cells (right), when the central density reaches $\rho_c=1.8\times10^{-11}$ g  cm$^{  -3}$. Scales are logarithmic and normalized to the rate of change of the total energy $d(E_\mathrm{k}+E_\mathrm{p}+U_\mathrm{i})/dt$.}.
\label{appA2}
\end{figure*}

\end{appendix}

\end{document}